\documentclass{aa} 
\usepackage{txfonts}
\usepackage{longtable}  
\usepackage{rotating}
\usepackage{natbib}
\usepackage{graphicx}
\usepackage{graphics}
\usepackage{psfrag}
\usepackage{amssymb}
\bibpunct{(}{)}{;}{a}{}{,}
\def\Teff{$T_{\mathrm{eff}}$}
\def\logg{\ensuremath{\log g}}
\def\vmic{$\upsilon_{\mathrm{mic}}$}

\def\vsini{\ensuremath{{\upsilon}\sin i}}
\def\kms{$\mathrm{km\,s}^{-1}$}

\def\exc{$\chi_{\mathrm{excit}}$}
\def\loggf{log$gf$}
\def\vr{${\upsilon}_{\mathrm{r}}$}

\def\espa{ESPaDOnS}

\def\nlte{non-LTE}
\def\llm{{\sc LLmodels}}
\def\hfs{{\it hfs}}

\def\width{{\sc WIDTH9}}
\def\synth{{\sc Synth3}}
\def\vald{{\sc VALD}}
\def\nist{{\sc NIST}}

\def\errvsini{$\sigma_{\ensuremath{{\upsilon}\sin i}}$}

\def\errvmic{$\sigma_{\upsilon_{\mathrm{mic}}}$}
\begin{document} 
\title{The chemical abundance analysis of normal early A- and late B-type 
stars\footnote{Tables \ref{line_abund} is only available in electronic form
at the CDS via anonymous ftp to cdsarc.u-strasbg.fr or via 
http://cdsweb.u-strasbg.fr/cgi-bin/qcat?J/A+A/.}} 
\subtitle{} 
\author{L. Fossati\inst{1}       \and
        T. Ryabchikova\inst{1,2} \and 
	S. Bagnulo\inst{3}	 \and
	E. Alecian\inst{4,5}	 \and
	J. Grunhut\inst{5}	 \and 
        O. Kochukhov\inst{6}     \and
	G. Wade\inst{5}
}
\institute{
	Institut f\"ur Astronomie, Universit\"{a}t Wien, 
	T\"{u}rkenschanstrasse 17, 1180 Wien, Austria.\\
	\email{fossati@astro.univie.ac.at,ryabchik@astro.univie.ac.at} 
	\and
	Institute of Astronomy, Russian Academy of Sciences, Pyatnitskaya 
	48, 119017 Moscow, Russia.\\
	\email{ryabchik@inasan.ru}
	\and
	Armagh Observatory, College Hill, Armagh BT61 9DG, Northern Ireland, 
	UK.\\
	\email{sba@arm.ac.uk} 
	\and 
	Observatoire de Paris-Meudon, LESIA, UMR 8111 du CNRS, 92195 Meudon 
	Cedex, France.\\
	\email{evelyne.alecian@obspm.fr}
        \and
	Physics Dept., Royal Military College of Canada, PO Box 17000, 
	Station Forces, K7K 4B4, Kingston, Canada.\\
	\email{Jason.Grunhut@rmc.ca,Gregg.Wade@rmc.ca}
        \and
	Department of Physics and Astronomy, Uppsala University, SE-751 20, 
        Uppsala, Sweden.\\
        \email{Oleg.Kochukhov@fysast.uu.se}
} 
\date{} 
\abstract
{Modern spectroscopy of early-type stars often aims at studying complex 
physical phenomena such as stellar pulsation, the peculiarity of the 
composition of the photosphere, chemical stratification, the presence of a 
magnetic field, and its interplay with the stellar atmosphere and the 
circumstellar environment. Comparatively less attention is paid to identifying 
and studying the "normal" A- and B-type stars and testing how the basic 
atomic parameters and standard spectral analysis allow one to fit the 
observations. By contrast, this kind of study is paramount eventually for 
allowing one to correctly quantify the impact of the various 
physical processes that occur inside the atmospheres of A- and B-type stars.}
{We wish to establish whether the chemical composition of the solar 
photosphere can be regarded as a reference for early A- and late B-type stars.}
{We have obtained optical high-resolution, high signal-to-noise
ratio spectra of three slowly rotating early-type stars (HD~145788,
21~Peg and $\pi$~Cet) that show no obvious sign of chemical
peculiarity, and performed a very accurate LTE abundance analysis of 
up to 38 ions of 26 elements (for 21~Peg), using a vast amount of spectral 
lines visible in the spectral region covered by our spectra.}
{We provide an exhaustive description of the abundance characteristics of the
three analysed stars with a critical review of the line parameters used to
derive the abundances. We compiled a table of atomic data for more than 1100
measured lines that may be used in the future as a reference. The
abundances we obtained for He, C, Al, S, V, Cr, Mn, Fe, Ni, Sr, Y, and Zr are 
compatible with the solar ones derived with recent 3D radiative-hydrodynamical
simulations of the solar photosphere. The abundances of the remaining studied 
elements show some degree of discrepancy compared to the solar photosphere. 
Those of N, Na, Mg, Si, Ca, Ti, and Nd may well be ascribed to non-LTE effects; 
for P, Cl, Sc and Co, non-LTE effects are totally unknown; O, Ne, Ar, and Ba 
show discrepancies that cannot be ascribed to non-LTE effects. The 
discrepancies obtained for O (in two stars) and Ne agree with 
very recent \nlte\ abundance analysis of early B-type stars in the solar 
neighbourhood.}
{}
\titlerunning{The abundance analysis for normal A- and B-type stars.}
\authorrunning{Fossati et al.}
\maketitle
\section{Introduction}\label{introduction}
In the last decade there has been dramatic improvement in the tools for the
analysis of optical stellar spectra, both from the observational and
theoretical perspective. New high-resolution echelle instruments have come 
online, which cover much broader spectral ranges than older single-order 
spectrographs. Data quality has also substantially improved in terms of 
signal-to-noise ratio (SNR), because of substantially greater instrument 
efficiency, and the use of large-size telescopes. Thanks to the vibrant 
observational activities of the past few years, and thanks to efficient and 
user-friendly data archive facilities, a huge high-quality spectroscopic 
database is now available to the public.

With the development of powerful and cheap computers, it has become practical 
to exploit these new data by performing spectral analysis using large spectral 
windows rather than selected spectral lines, at a level of realism heretofore 
impossible. The high accuracy of observations and modelling techniques now 
allows, for example, stretching the realm of abundance analysis to faster 
rotators than was possible in the past, but also provides the possibility of 
learning more about the structure of the stellar atmospheres and the ongoing 
physical processes, especially when spectral synthesis {\it fails} to 
reproduce the observations. For instance, observed discrepancies between 
observed and synthetic spectra have allowed us to discover that the signature 
of chemical stratification is ubiquitous in the spectra of some chemically 
peculiar stars \citep{Bagetal01,wade2001,RWL2003} and to perform accurate 
modelling of this stratification in the atmospheres of Ap stars 
\citep[for instance,][]{RLK05,KTRM06}.

However, our inability to reproduce observations frequently stems for a very 
simple cause: that atomic data for individual spectral lines are incorrect. 
For solar type stars it is possible to construct a "reference" list of 
reliable spectral lines with reliable atomic data through the comparison of
synthetic spectra with the solar observed spectrum, since the solar abundances 
are accurately known. In many cases, however, the solar spectrum cannot 
provide the required information, because the temperature of the target stars 
is significantly different from that of the sun. This problem can be overcome 
by adopting an analogous reference at a temperature reasonably close to the 
target temperature. The method involves a selection of a set of suitable 
reference stars for which very high quality spectra are available. Then  
an accurate determination of the stellar photospheric parameters and an 
accurate abundance analysis are performed with the largest possible number of 
spectral lines and the best possible atomic data. Finally, those spectral 
lines exhibiting the largest discrepancies from the model fit are identified, 
and their atomic data revised by assuming that the average abundance (inferred 
from the complete sample of spectral lines of that element) is the correct 
one. In this process it is important to take effects into account that can 
potentially play a significant role in all stellar atmospheres, such as 
variations in the model structure from non-solar abundances and \nlte\ 
effects.

In this paper we address the problem of establishing references for effective
temperatures around 10000--13000\,K. This is in some respect the easiest 
temperature range to study, as well as one of the most interesting. This 
temperature is close to ideal because the spectra of stars in this interval 
are generally unaffected by severe blending. It is also relevant 
because stars in this temperature range display spectroscopic 
peculiarities (chemical abundance peculiarities, stratification, Zeeman 
effect, etc.) that reflect physical conditions and processes of interest 
for detailed investigation. A crucial prerequisite for studying and 
characterising these phenomena is the capacity to model the underlying 
stellar spectrum in detail, and this requires high quality atomic data. 

The highest degree of accuracy in abundance analysis is reached for
sharp-lined stars. Unfortunately, these objects are quite rare among
A- and B-type stars, which are generally characterised by high
rotational velocities. Furthermore, most of the slowly rotating stars in
the chosen temperature region belong to various groups of magnetic and
non-magnetic, chemically peculiar objects. As a matter of fact, many
previous studies aimed at determining the chemical composition of
"normal" early A- and late B-type stars were based on samples
"polluted" by moderately chemically peculiar stars. For instance,
the work by \citet{HH2003} includes the sharp-lined HgMn star 53 Aur.
\citet{HL1993} searched for compositional differences among A-type
stars, but four out of six programme stars are in fact
classified as hot Am stars on the basis of the abundances of the heavy
elements Sr-Y-Zr-Ba, which are considered as diffusion indicators
\citep[i.e., Sirius, $o$~Peg, and $\theta$~Leo, see ][]{HH2003}. The 
complexity of the problem of distinguishing
between normal A and marginal Am stars is further stressed by
\citet{AU2007}. The aim of the present paper is to search for
sharp-lined early A- or late B-type stars with a chemical composition
as close as possible to the solar one. As a final outcome, one could
assess whether the chemical composition of the solar photosphere may be
considered at least in principle as a \textit{reference} for the
composition of the early A- and late B-type stars. If such a star is
found, this will not imply that the solar composition is the most
characteristic for the slowly rotating A- and B-type stars, but will be
used as further evidence that any departure from the composition of
the solar photosphere has to be explained in terms of diffusion or
other physical mechanisms that are not active at the same efficiency
level in the solar photosphere.

Our work is based on a very detailed and accurate study of a vast
sample of spectral lines. As a by-product, we provide a list of more
than 1100 spectral lines from which we have assessed the accuracy of
the corresponding atomic data. Such a list may serve as a future reference 
for further abundance analysis studies of stars with a similar
spectral type.

This paper is organised as follows. Sect.~\ref{observations} describes the
target selection, observations and data reduction, 
Sect.~\ref{fundamental parameters} presents methods and results for the choice 
of the best fundamental parameters that describe the atmospheres of the 
programme stars, Sect.~\ref{abundance analysis} presents the methods and 
results of the abundance analysis of the programme stars. Our results are 
finally discussed in Sect.~\ref{discussion}.
\section{Target selection, observations, and data reduction}\label{observations}
For our analysis we need a late A-type or early B-type star with a sharp-line 
spectrum (hence the star must have a small \vsini) and one exhibiting the 
least possible complications due to phenomena such as non-homogeneous surface
distribution of chemical elements, pulsation, or a magnetic 
field. Finding such a target is not a simple task, because most of the 
early-type stars are fast rotators. Slowly rotating A- and B-type stars 
generally show some type of chemical peculiarity often associated to the 
presence of abundance patches, a magnetic field (which broadens, or even splits 
spectral lines), and chemical stratification. Even Vega, which has been 
considered for a long time as the prototype of a "normal", slow rotating 
A-type star, is in fact currently classified as a $\lambda$~Boo star and
discovered to actually be a fast rotating star seen pole-on, exhibiting, as 
such, distorted line profiles \citep{adelman1990,yoon}. Based on our 
knowledge, we reached the conclusion that the most suitable target for our 
project is the B9 star 21~Peg (HD~209459), which is known from previous 
studies as a "normal" single star with $\vsini \sim 4$\,\kms\ 
\citep[see, e.g.][]{sadakane81}.

We felt it was necessary to consider additional targets of our spectral 
analysis, for two main reasons. Since we intend to provide an accurate
reference for the typical abundances of the chemical elements in A-
and B-type stars (and compare these values with the solar ones), we
need to cross-check with further examples whether the results obtained
for 21~Peg are similar to those of other ''normal", slow rotating
A-type stars. Second, to check the accuracy of the astrophysical measurements
of the \loggf\ values, which is a natural complement of the present
work. Both these goals are best achieved
with the use of abundance values that have been obtained with a
homogeneous method, rather than from a mixed collection of data from
the literature. Therefore, we have also analysed another two stars of
similar temperature as 21~Peg, i.e., HD~145788 (HR~6041) and 
$\pi$~Cet (HR~811). Both stars fulfill our requirements, although are 
slightly less ideal than 21~Peg. HD~145788, suggested to us by Prof.~Fekel, 
is a slowly rotating single star with \vsini\ $\sim$ 8\,\kms \citep{fekel}. 
$\pi$~Cet, a SB1 with $\vsini \sim 20$\,\kms, shows an infrared excess, and is
a suspected Herbig Ae/Be star \citep{malfait}. Since its spectrum is not 
visibly contaminated by the companion, it still serves our purpose. 
$\pi$~Cet was also already used as a normal comparison star in the abundance 
study of chemically peculiar stars by \citet{smith93}. 

The star 21~Peg was observed five times during two observing nights in August
2007, with the FIES instrument of the North Optical Telescope (NOT).
FIES is a cross-dispersed high-resolution \'{e}chelle spectrograph
that offers a maximum spectral resolution of R = 65\,000, covering
the entire spectral range 3700--7400\,\AA.

Data were reduced using a pipeline developed by D.~Lyashko, which is
based on the one described by \citet{dima}. All bias and flat-field
images were median-averaged before calibration, and the scattered
light was subtracted by using a 2D background approximation. For
cleaning cosmic ray hits, an algorithm that compares the direct and
reversed observed spectral profiles was adopted. To determine the 
boundaries of echelle orders, the code uses a special template for each 
order position in each row across the dispersion axis. The shift of the 
row spectra relative to the template was derived by a cross-correlation
technique. Wavelength calibration of each image was based on a single
ThAr exposure, recorded immediately after the respective stellar time
series, and calibrated by a 2D approximation of the dispersion
surface. An internal accuracy of $\sim 100$\,ms$^{-1}$ was
achieved by using several hundred ThAr lines in every echelle order.

Each reduced spectrum has a SNR per pixel of about 300
at 5000\,\AA.  All five spectra are fully consistent among themselves, 
which confirms that the star is not variable. This allowed us to combine 
all data in a unique spectrum with a final SNR of about 700.

Because of the very low \vsini\ of 21~Peg, we made use of a very 
high-resolution spectrum ($R=120\,000$) obtained with the Gecko instrument 
(now decommissioned) of the Canada-France-Hawaii Telescope 
\citep[CFHT,][]{john1998} to measure this parameter. The spectrum covers the 
ranges 4612--4640\,\AA\ and 5160--5192\,\AA, which are too short to perform 
a full spectral analysis, but sufficient to measure \vsini\ with high accuracy.
According to \citet{hubrig06} the mean longitudinal magnetic field of 
21~Peg is $-$144$\pm$60\,G, which excludes the possibility that the star 
has a structured magnetic field.

The spectrum of HD~145788 was obtained with the cross-dispersed
\'{e}chelle spectrograph HARPS instrument attached at the 3.6-m ESO La
Silla telescope, with an exposure time of 120\,s, and reduced with the
online pipeline
\footnote{\tt
http://www.ls.eso.org/lasilla/sciops/3p6/harps/\\software.html\#pipe}.
The reduced spectrum has a resolution of 115\,000, and a SNR per pixel of 
about 200 at 5000\,\AA.  The spectral range is 3780--6910\,\AA\ with a 
gap between 5300\,\AA\ and 5330\,\AA, because one echelle order is lost in 
the gap between the two chips of the CCD mosaic detector.

The star $\pi$~Cet was observed with the \espa\ instrument of the CHFT
February, 20 and 21 2005. \espa\ consists of a table-top,
cross-dispersed echelle spectrograph fed via a double optical fiber
directly from a Cassegrain-mounted polarisation analysis module. Both
Stokes $I$ and $V$ spectra were obtained throughout the spectral range
3700 to 10400\,\AA\ at a resolution of about 65\,000. The spectra
were reduced using the Libre-ESpRIT reduction package 
\citet[][ and in prep.]{donatietal1997}. The two spectra (each obtained 
from the combination of four 120\,s sub-exposures) were combined into a final
spectrum that has an SNR per pixel of about 1200 at 5000\,\AA.

The observation of $\pi$~Cet enters in the context of a large
spectropolarimetric survey of Herbig Ae/Be stars. Least-squares
deconvolution \citep[LSD,][]{donatietal1997} was applied to the spectra of
$\pi$~Cet assuming a solar abundance line mask corresponding to an
effective temperature of 13000\,K. The resulting LSD profiles show a
clean, relatively sharp mean Stokes $I$ profile, corresponding to
$\vsini = 20 \pm 1$\,\kms, and no detection of any Stokes $V$ signature
indicative of a photospheric magnetic
field. Integration of Stokes $V$ across the line using Eq.~(1) of
\citet{wade2000} yields longitudinal magnetic fields consistent with zero
field and with formal 1$\sigma$ uncertainties of about 10\,G. The
high-resolution spectropolarimetric measurements therefore provide no
evidence of magnetic fields in the photospheric layers of the star.

\begin{figure}[ht]
\begin{center}
\includegraphics[width=90mm]{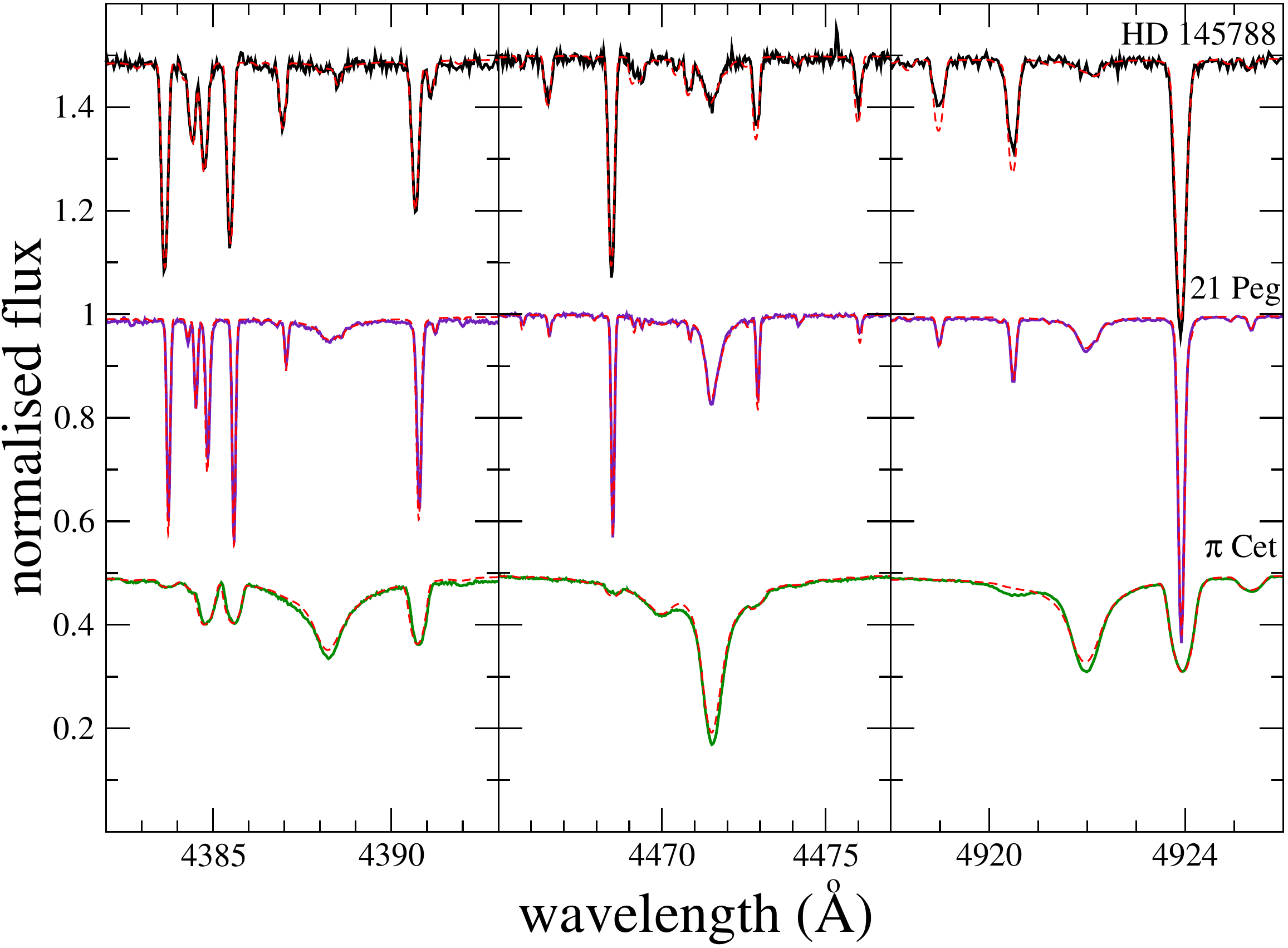}
\caption{Samples of the spectra of HD~145788, 21~Peg and $\pi$~Cet around
three \ion{He}{i} lines: 4387\,\AA, 4471\,\AA\ and 4921\,\AA\ in comparison 
with our final synthetic profiles (dashed lines) calculated for each line. 
The spectra of HD~145788 and $\pi$~Cet are shifted upwards and downwards 
of 0.5, respectively.}
\label{portions}
\end{center}
\end{figure}

All the spectra of the three stars were normalised by fitting a spline
to carefully selected continuum points. For each object, radial
velocities \vr\ were determined by computing the median of the results
obtained by fitting synthetic line profiles of several individual 
carefully selected lines into the observed spectrum. The \vr\ values are 
listed in Table~\ref{parameters}, and their uncertainty is of the order of
0.5\,\kms. The spectra were then shifted to the wavelength rest
frame. Selected spectral windows containing the observed blue \ion{He}{i} 
lines, together with the synthetic profiles, are displayed in 
Fig.~\ref{portions}.
%
\section{Fundamental parameters}\label{fundamental parameters}
Fundamental parameters for the atmospheric models were obtained using
photometric indicators as a first guess. For their refined estimate,
we performed a spectroscopic analysis of hydrogen lines and metal lines, and
as a final step, compared the observed and computed energy 
distributions. The spectroscopic tuning of the fundamental
parameters is needed since different photometries and calibrations would give
different parameters and uncertainties. The spectroscopic analysis will provide
a set of parameters that fit all the parameter indicators consistently,
with less uncertainties. Model atmospheres were calculated 
with \llm, an LTE code that uses direct sampling of the line opacities 
\citep{denis2004} and allows computing models with an individualised 
abundance pattern. Atomic parameters of spectral lines used for model 
atmosphere calculations were extracted from the \vald\ database 
\citep{vald1,vald2,vald3}.

Before applying the spectroscopic method, we estimated the star's \vsini. 
For 21~Peg, a \vsini\ value of $3.76 \pm 0.35$\,\kms\ was derived from the 
fit with a synthetic spectrum to 21 carefully selected lines observed with 
the Gecko instrument. This value agrees very well with the 3.9\,\kms\ value 
derived by \citet{john1998}. Achieving such high precision was possible thanks
to the high quality of the spectrum and the low \vsini.  The \vsini\ 
values for HD~145788 and $\pi$~Cet, $10.0 \pm 0.5$\,\kms\ and
$20.2 \pm 0.9$\,\kms, respectively, were derived from fitting about 20 
well-selected lines along the whole available spectral region.

In the next sections, we describe the determination of the atmospheric 
parameters: \Teff, effective gravity (\logg), and
microturbulent velocity (\vmic), and their uncertainties. 
The fundamental parameters finally adopted for 21~Peg, 
HD~145788, and $\pi$~Cet are given in Table~\ref{parameters}.
\begin{table*}[ht]
\caption[ ]{Adopted atmospheric parameters for the analysed stars.}
\label{parameters}
\begin{center}                     
\begin{tabular}{cccccccc}
\hline
\hline
Star & \Teff & \logg  & \vmic  & \errvmic  & \vsini & \errvsini & \vr    \\
Name &   [K] & [cgs]  & [\kms] & [\kms]    & [\kms] & [\kms]    & [\kms] \\
\hline											
21~Peg    & 10400 & 3.55 & 0.5 & 0.4 & 3.76 & 0.35 &  0.5    \\
HD~145788 &  9750 & 3.70 & 1.3 & 0.2 & 10.0 & 0.5  & $-$13.9 \\
$\pi$~Cet & 12800 & 3.75 & 1.0 & 0.5 & 20.2 & 0.9  &  12.5   \\
\hline
\end{tabular}
\end{center}                     
\smallskip 

The uncertainties on \Teff, \logg, and \vr\ are 200\,K, 0.1\,dex, and 0.5\,\kms, respectively.
\end{table*}

%
\subsection{Photometric indicators}
Since none of the three programme stars, HD~145788, 21~Peg or
$\pi$~Cet are known to be photometrically variable or peculiar, we can
use temperature and gravity calibrations of different photometric
indices for normal stars to get a preliminary estimate of the
atmospheric parameters. The effective temperature (\Teff) and gravity
(\logg) were derived from Str\"omgren photometry \citep{hauck} with
calibrations by \citet{moon1985} and by \citet{napiwotzki1993},
and from Geneva photometry \citep{rufener} with the calibration by
\citet{north1990}.
The mean parameters from the three calibrations that were used as starting 
models are the following ones: \Teff\ = 9675$\pm$75\,K, 
\logg\ = 3.72$\pm$0.03 for HD~145788; \Teff\ = 10255$\pm$115\,K, 
\logg\ = 3.51 for 21~Peg; \Teff\ = 13200$\pm$65\,K 
\logg\ = 3.77$\pm$0.15 for $\pi$~Cet. No error bar is given for the \logg\ 
of 21~Peg since all three calibrations give the same value.
\subsection{Spectroscopic indicators}
\subsubsection{Hydrogen lines}\label{hlines}
For a fully consistent abundance analysis, the photometric parameters have to 
be checked and eventually tuned according to spectroscopic indicators, such as 
hydrogen line profiles. In the temperature range where HD~145788, 21~Peg, and
$\pi$~Cet lie, the hydrogen line wings are less sensitive to \Teff\ 
than to \logg\ variations, but temperature effects can still be visible in the 
part of the wings close to the line core. For this reason hydrogen lines are 
very important not only for our analysis, but in general for every consistent 
parameter determination. To spectroscopically derive the fundamental parameters 
from hydrogen lines, we fitted synthetic line profiles, calculated with 
\synth\ \citep{synth3}, to the observed profiles. \synth\ incorporates 
the code by 
\citet{barklem2000}\footnote{\tt http://www.astro.uu.se/$\sim$barklem/hlinop.html} 
that takes into account not only self-broadening but also Stark 
broadening (see their Sect.~3). For the latter, the default mode of \synth,
adopted in this work, uses an improved and extended HLINOP routine 
\citep{kurucz93}.

\begin{figure*}[ht]
\sidecaption
\includegraphics[width=130mm]{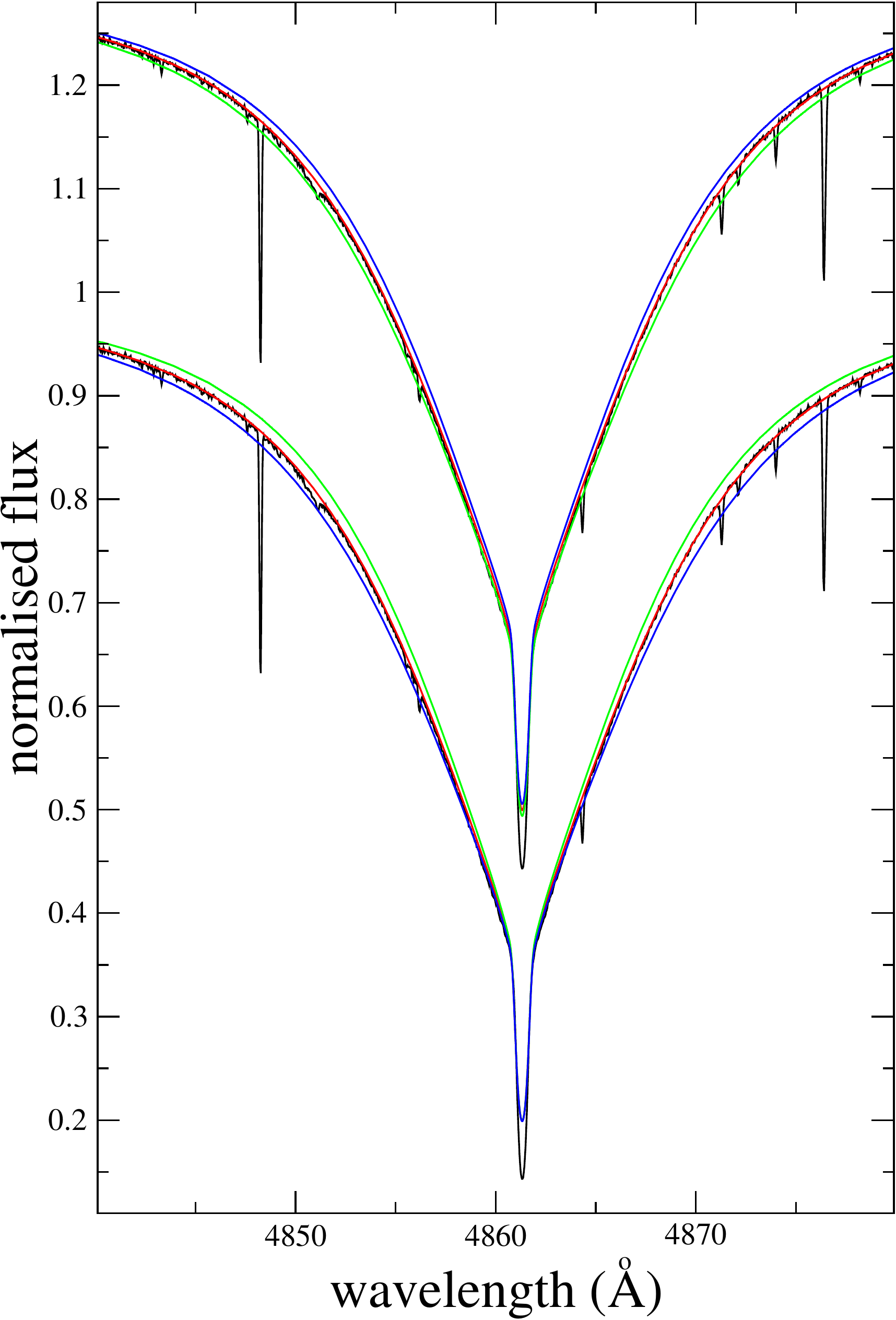}
\caption{Observed H$\beta$ line profile for 21~Peg (black solid line), 
compared to synthetic profiles (red, blue, and green lines). The red solid 
lines were obtained assuming the best \Teff\ and \logg\ values of 
Table~\ref{parameters}. The blue lines show the synthetic profile by increasing
\Teff\ by 200\,K (top spectrum) or \logg\ by 0.1\,dex (bottom spectrum). The
green lines show the synthetic profile by decreasing \Teff\ by 200\,K (top 
spectrum) or \logg\ by 0.1\,dex (bottom spectrum).} 
\label{hydrogen_21peg_hbeta} 
\end{figure*}

Figure~\ref{hydrogen_21peg_hbeta} shows the comparison between the observed 
H$\beta$ line profile for 21~Peg and the synthetic profiles calculated with 
the adopted stellar parameters. In Fig.~\ref{hydrogen_21peg_hbeta} we also  
added the synthetic line profiles calculated with 1$\sigma$ error bars on 
\Teff\  ($\pm$ 200\,K; upper profile) and \logg\  ($\pm$ 0.1\,dex; lower 
profile). The same profiles with the same uncertainties, but for H$\gamma$, 
H$\beta$, and H$\alpha$ (from left to right) for the three programme stars, are 
shown in Figs.~\ref{hydrogen_hd145788}, \ref{hydrogen_21peg}, and 
\ref{hydrogen_hd17081} (online material). 

The three hydrogen lines (H$\gamma$, H$\beta$, and H$\alpha$) used to 
spectroscopically improve the fundamental parameters for HD~145788 gave 
slightly different results both for \Teff\ and for \logg. As final values, we 
took their mean (\Teff\ = 9750\,K; \logg\ = 3.7). This discrepancy is 
visible in Fig.~\ref{hydrogen_hd145788} (online material), but it lies within 
the errors given for \Teff\ and \logg. The spectrum of HD~145788 also allowed  
a good normalisation of the region bluer than H$\gamma$. We synthesised this 
region to check the quality of our final fundamental parameters finding 
a very good fit for the three hydrogen lines H$\delta$, H$\epsilon$, and H$_8$.

For 21~Peg we obtained the same temperature estimates from H$\alpha$ and
H$\beta$ (\Teff\ = 10400\,K) and by 200\,K less from the fitting of H$\gamma$.
We adopted a final \Teff\ of 10400\,K. To fit all three hydrogen lines, we need 
slightly different values of \logg: 3.47 for H$\gamma$, 3.54 for 
H$\beta$ and 3.57 for H$\alpha$. We finally adopted \logg\ = 3.55 
taking possible continuum normalisation problems into account, in particular, 
for H$\gamma$.

The temperature determination for $\pi$~Cet was more difficult thanks to the
weak effect that this parameter has on the hydrogen lines at about 13000\,K and
to the slightly peculiar shape of the H$\alpha$ line. As explained in
Sect.~\ref{discussion}, $\pi$~Cet probably shows a small emission signature 
in the region around the core of H$\alpha$ possibly because of a 
circumstellar disk. This region is the one where 
\Teff\ effects are visible, making it almost impossible to obtain a good 
temperature determination from this line. H$\gamma$ and H$\beta$ gave best 
temperatures of 12700\,K and 12900\,K, respectively, leading to a final adopted 
value of 12800\,K. Confirmation of this value was then given by the 
spectrophotometry (see Sect.~\ref{spectrophot}). The results of LTE abundance 
analysis (Sect.~\ref{abundance analysis}) show that $\pi$~Cet has little He 
overabundance that 
leads to an overestimation of \logg\ if the He abundance is not taken into 
account in the model atmosphere calculation \citep{auer}. For this 
reason we derived the first set of fundamental parameters (\Teff\ = 12800\,K; 
\logg\ = 3.80) and then the He abundance 
($\log(N_{\rm He}/N_{\rm tot})$ = $-$0.97\,dex). As a next step we recalculated 
a set of model atmospheres with the derived He abundance and re-fit the 
hydrogen line profiles. The best-fit gave us the same temperature, but weak 
effective gravity (\logg\ = 3.75). As expected, the He overabundance 
is acting as pressure, requiring an adjustment of \logg\ to be balanced. 
We obtained the He abundance from the fitting of the blue He line wings.
The blue He lines are, in general, considered as showing very little \nlte\ 
effect \citep{LL98}, and as we only used the line wings, this leads us to 
conclude that our results should not be affected by \nlte\ effects and that 
the He is overabundant in $\pi$~Cet. The best fit to the blue \ion{He}{i} 
lines of $\pi$~Cet is shown in Fig.~\ref{portions}.

The example of $\pi$~Cet is important because it clearly shows the effect of 
the single element abundance on the parameter determination, not only for 
chemically peculiar stars (for which this effect is well known and often, 
but not always, taken into account), but also for chemically "normal" stars.

The set of parameters that best fit the hydrogen line profiles could not 
be unique. For 21~Peg, we checked that using a combination of a lower 
temperature and lower gravity or else higher temperature and higher gravity 
increases the standard deviation of the fit of the H$\beta$ line wings by 
$\sim$25\%. The result is that a different combination of \Teff\ 
and \logg\ could in principle provide a similar fit. For this reason the 
derived fundamental parameters should be checked with other indicators, such 
as the analysis of metallic lines (ionisation and excitation equilibrium) and 
the fitting of the spectral energy distribution. The latter is more 
important because ionisation and excitation equilibrium should be strictly 
used only with a full \nlte\ treatment of the line formation.
\subsubsection{Metallic lines}\label{metal_lines}
The metallic-line spectrum may also provide constraints on the atmospheric 
parameters. If no deviation from the local thermodynamic equilibrium 
(LTE) is expected, the minimisation of the correlation between individual line 
abundances and excitation potential, for a certain element/ion, allows one to 
check the determined \Teff. Then the balance between different ionisation 
stages of the same element provides a check for \logg. The microturbulent 
velocity, \vmic\ is then determined by minimising the correlation between 
individual abundances and equivalent widths for a certain element. 
Determining the fundamental parameters performed in this way 
has to be done iteratively since, for example, a variation in \Teff\ leads 
to a change in the best \logg\ and \vmic.
\begin{figure}[ht]
\begin{center}
\includegraphics[width=90mm]{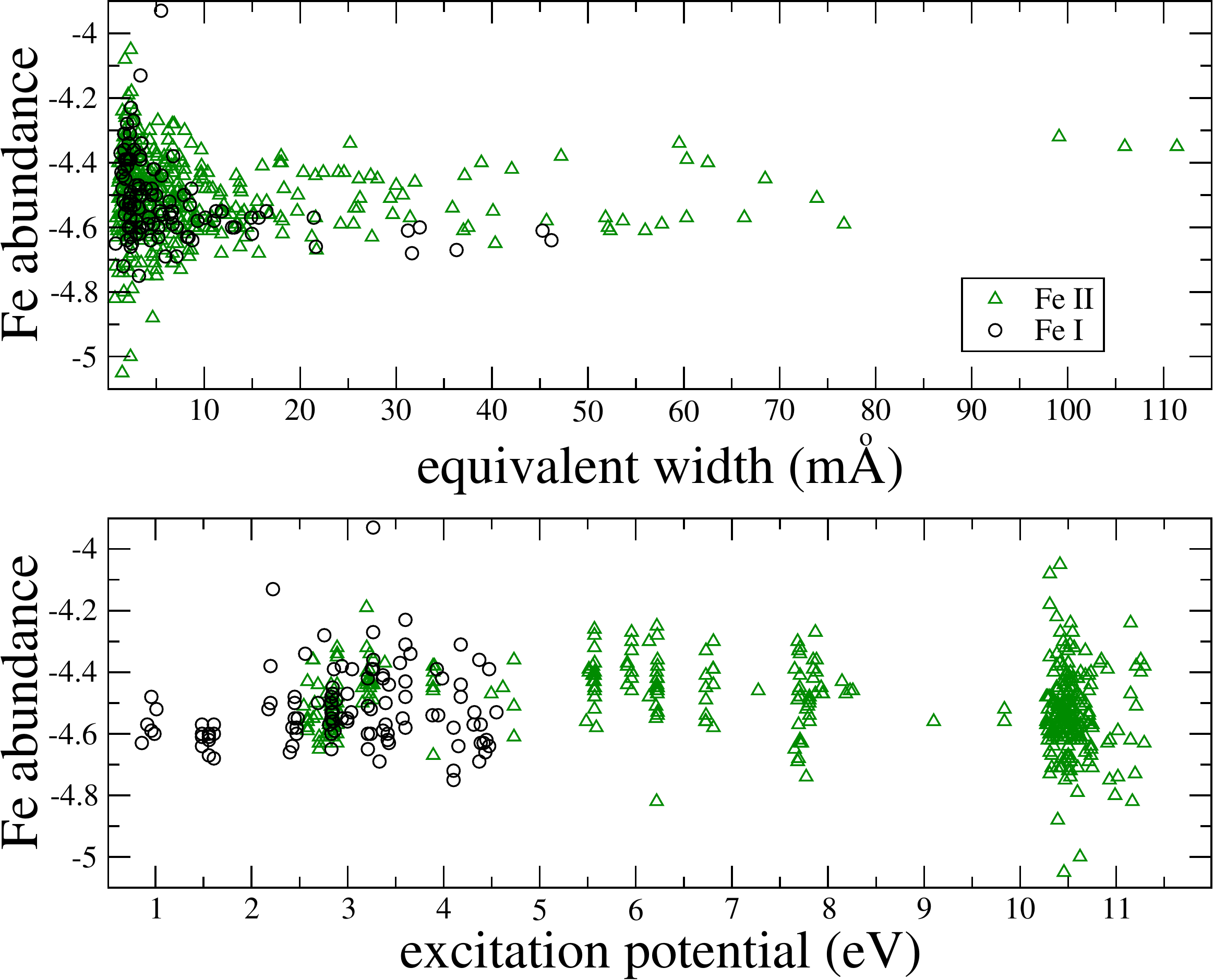}
\caption{Iron abundance vs. equivalent widths (upper panel) and excitation
potential (lower panel) for 21~Peg. The open circles indicate 
\ion{Fe}{i}, while the open triangles indicate \ion{Fe}{ii}.} 
\label{metallic lines} 
\end{center} 
\end{figure}

Figure~\ref{metallic lines} shows the correlations of \ion{Fe}{i} and 
\ion{Fe}{ii} abundances with the equivalent widths (upper panel) and with the 
excitation potential (lower panel) for 21~Peg. The correlation with the 
equivalent widths is shown for a \vmic\ of 0.5\,\kms, which we infer to be the 
best value for 21~Peg, since the slope of the linear fit for \ion{Fe}{i} is 
$-4.897\times10^{-3} \pm 1.560\times10^{-3}$\,m\AA$^{-1}$ and for \ion{Fe}{ii} 
is $2.146\times10^{-5} \pm 4.058\times10^{-4}$\,m\AA$^{-1}$. Here we gave a 
preference to the result obtained from \ion{Fe}{ii} because of the higher 
number of measured \ion{Fe}{ii} lines in a wider range of equivalent
widths. The same analysis was made for HD~145788 and for $\pi$~Cet. 
The error bar on \vmic\ was calculated using the error bar of the slope
(abundance vs. equivalent width) derived from a set of different \vmic. The 
uncertainties listed in Table~\ref{parameters} are given considering
2$\sigma$ on the error bar of the derived slopes. Considering a 1$\sigma$ error
bar, the uncertainties on \vmic\ are of 0.1\,\kms for HD~145788 and 21~Peg and 
of 0.2\,\kms\ for $\pi$~Cet. 

According to previous works, deviation from LTE of the \ion{Fe}{ii} 
lines is expected to be very small 
\citep[$\sim$ 0.02\,dex ][]{gigas1986,rentzsch1996} for the analysed stars, 
such that the absence of any clear correlation between the line \ion{Fe}{ii} 
abundance and the excitation potential confirms the \Teff\ derived from the 
hydrogen lines. 

For the \ion{Fe}{i} lines, deviations from LTE of about $+$0.3\,dex are given 
by \citet{gigas1986} and \citet{rentzsch1996}. However, both \citet{gigas1986} 
and \citet{rentzsch1996}, as well as \citet{HH2003}, used the same model atom, 
which includes 79 \ion{Fe}{i} and 20 \ion{Fe}{ii} energy levels. We note that 
the highest energy level in their model atom for \ion{Fe}{ii} has an excitation 
energy of about 6\,eV, while the ionisation potential is 16.17\,eV. Such a 
model atom does not provide collisional coupling of \ion{Fe}{ii} to 
\ion{Fe}{iii}, which operates for the majority of iron atoms in line formation 
layers below $\log \tau_{5000} =-1$. Unfortunately, the existing NLTE 
calculations for Fe are not accurate enough to be applied now to our 
stars. Clearly, an extended energy-level model atom is needed for a reliable 
\nlte\ analysis of Fe. The ionisation equilibrium for different elements/ions 
(or its violation) can be seen in Table~\ref{abundance} and is discussed in 
Sect.~\ref{abundance analysis}. 
\subsection{Spectrophotometry}\label{spectrophot}
\begin{figure*}[ht]
\sidecaption
\includegraphics[width=140mm]{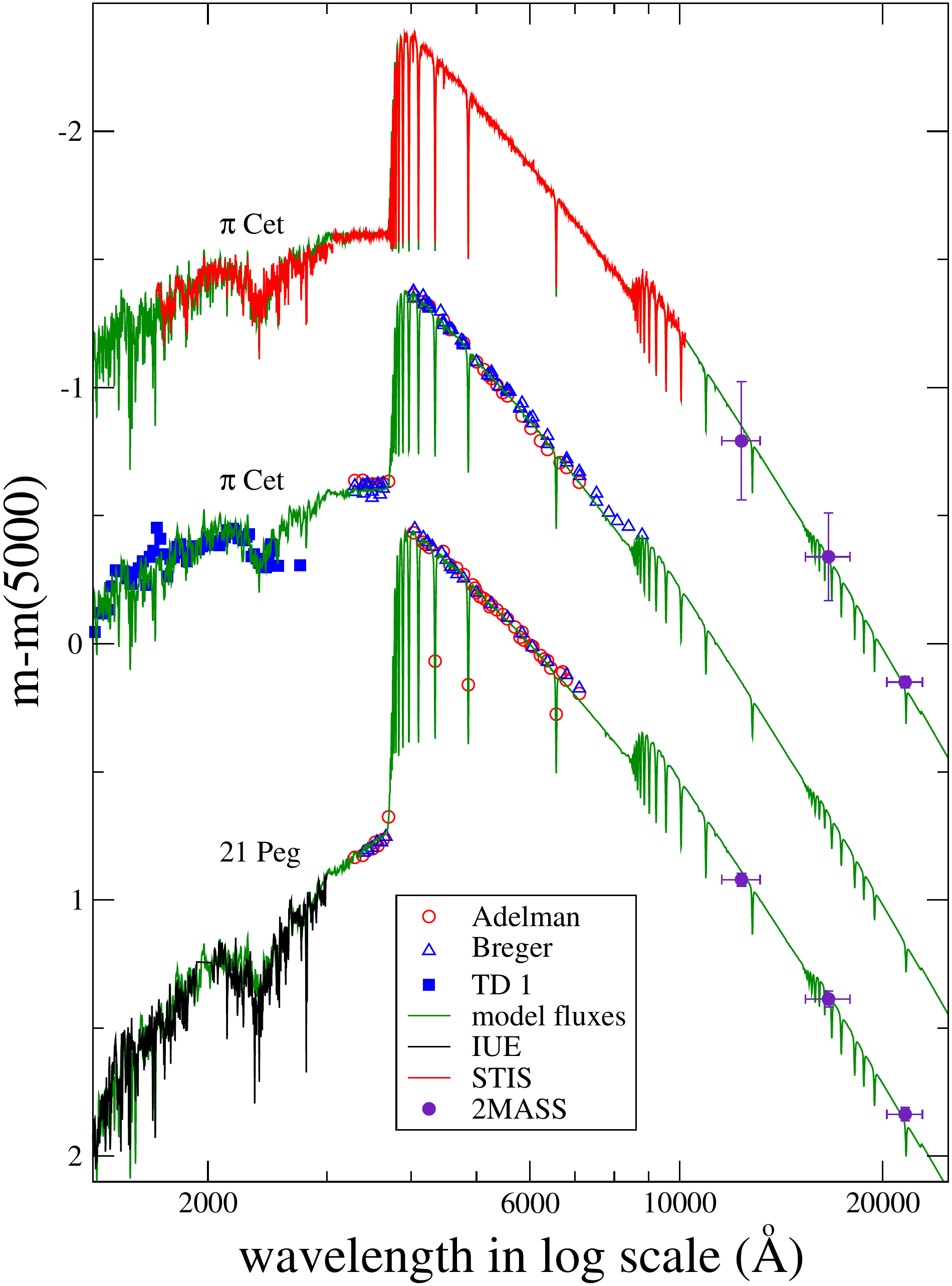}
\caption{Comparison between \llm\ theoretical fluxes  
calculated with the fundamental parameters and abundances derived for 
21~Peg and $\pi$~Cet (green solid line) with spectrophotometries by 
\citet{adelman1989} (open red circles) and \citet{breger} 
(open blue triangles), IUE calibrated fluxes (full black line), TD~1 
observations (full blue squares) and STIS spectrum (full red line). The 
violet full dots represent the 2MASS photometry for which the given error 
bar in wavelength shows the wavelength range covered by each filter. The 
model fluxes for 21~Peg and $\pi$~Cet, shown in the middle of the panel, 
were convolved to have the same spectral resolution of the IUE fluxes 
({\it R}~$\sim$~900), while the upper of $\pi$~Cet has a spectral 
resolution 700, approximately the same of STIS spectra. For a better
visualisation of the whole spectral region the x-axis is on a logarithmic 
scale.}
\label{sph_paper} 
\end{figure*}

For a complete self-consistent analysis of any star, one should reproduce 
the observed spectral energy distribution with the adopted parameters for a
model atmosphere. In the optical spectral region, spectrophotometry was 
only available for 21~Peg and $\pi$~Cet, while in the ultraviolet, IUE spectra 
were available for all three stars. For $\pi$~Cet ultraviolet spectrophotometry 
from the TD-1 satellite \citep{TD1} was also available, along with 
the flux calibrated spectra from STIS at HST \citep{hst3}. The comparison 
between the observed flux distributions and the model fluxes calculated 
with the adopted atmospheric parameters for 21~Peg and $\pi$~Cet is shown in 
Fig.~\ref{sph_paper}. For HD~145788 we estimated a reddening 
E(B-V)$\approx$0.2 from the dust maps of \citet{schlegel}. The comparison of 
reddened model fluxes with the available IUE spectrum, Johnson UBV photometry 
\citep{nicolet78}, Geneva 
photometry\footnote{\tt http://obswww.unige.ch/gcpd/ph13.html} and 2MASS
photometry \citep{cutri} is shown in Fig.~\ref{sph_hd145788} (online material). 
This plots supports the value of E(B-V)=0.2 and shows good agreement 
between all the observations and the model fluxes, confirming 
the obtained fundamental parameters, and also the importance of considering 
reddening in the analysis of relatively nearby stars such as HD~145788.

The optical spectrophotometry was taken from 
\citet{adelman1989} and \citet{breger}. All flux measurements were 
normalised to the flux value at 5000\,\AA\ obtained from observed data 
for $\pi$~Cet and 21~Peg and from the model fluxes for HD~145788.
Also for 21~Peg and $\pi$~Cet, we extended the comparison to the near infrared 
region including 2MASS photometry \citep{cutri}.
Following \citet{netopil}, we obtained a reddening 
E(B-V)$\leq0.01\pm0.01$\,mag for both $\pi$~Cet and 21~Peg. A good agreement 
between the unreddened model fluxes and the observed spectrophotometry from 
UV to near infrared confirms negligible reddening for these two stars.

The agreement between the IUE observation and the model fluxes is very good
for HD~145788, but the IUE data alone cannot be used as a check of \Teff, 
since a variation of about 1000--1500\,K is needed to make any visible 
discrepancy. At the \Teff\ of HD~145788 hydrogen lines were still quite
sensitive to temperature variations, so we did not need accurate
spectrophotometry for a reliable effective temperature estimate. The 
slight disagreement between the model fluxes and the 2MASS photometry could 
stem from our not accounting for reddening in the calculation of the model 
fluxes since the derived value is too uncertain.

The model fluxes of 21~Peg are in good agreement with the observations 
from the ultraviolet up to the near infrared. For 21~Peg and $\pi$~Cet, few 
data points in the red appear above the model fluxes. This often
happens with the optical spectrophotometry as shown in Fig.~1 of 
\citet{adelman02}. We do not have a definite explanation for this effect,
although taking a very accurate measurement of the reddening into account 
could remove this inconsistency. As for the hydrogen lines, we 
checked the fit of the spectral energy distribution to model fluxes 
calculated with a combination of a lower temperature and lower gravity or 
a higher temperature and higher gravity. In both cases the fit clearly becomes  
worse mainly in the spectral region around the Balmer jump. This definitively 
excludes the possibility that a combination of fundamental parameters, which 
gives a worse but still reasonable fit to the hydrogen line profiles, should 
considered.

For $\pi$~Cet we obtained several spectrophotometric observations from 
different sources. In particular, the data in the optical region around the 
Balmer jump were conclusive for \Teff\ determination, as mentioned already in 
Sect.~\ref{hlines}. The STIS spectrum  shown in Fig.~\ref{sph_paper} is formed 
by three separated spectra (UV, visible, IR) with about the same spectral 
resolution. We noticed a small (but not negligible) vertical jump at the 
overlapping wavelengths between the three spectra. For this reason we decided 
to fit them separately. This decision was based on an adjustment 
performed at the level of the flux calibration \citep{hst1,hst2}. 

The use of an energy distribution plays a crucial role in adjusting the 
atmospheric parameters because it is independent of the photometric 
calibrations (different for each author), of the hydrogen line fitting 
(reduction and normalisation dependent), and of the ionisation equilibria 
approach (dependent on several effects such as the adopted atomic line data 
and \nlte\ effects). 

The uncertainties on \Teff\ and \logg\ were derived from the hydrogen
line fitting. This way of deriving the error bars on the parameters also 
includes the SNR of the observations. For all three stars, we
derived an error bar on the \Teff\ of 200\,K and 0.1\,dex in \logg, as also 
shown in Figs.~\ref{hydrogen_21peg}, \ref{hydrogen_hd145788}, and 
\ref{hydrogen_hd17081} (online material).
\subsection{Comparison with previous determinations}\label{comp_param}
Only one previous temperature determination for HD~145788 is
known. \citet{glago} determined a \Teff\ of 9600\,K from the reddening free 
$Q$ parameter \citep{johnson1953} and of 9100\,K from the $X$ parameter derived 
from multicolour photometry. The \Teff\ derived from the $Q$ parameter agrees 
with our estimation.

The star 21~Peg was analysed for abundances and atmospheric parameters several 
times in the past. The atmospheric parameters extracted from literature are
collected in Table~\ref{comp_param_21peg}, together with the methods of their 
determination. The last column of Table~\ref{comp_param_21peg} lists the 
methods adopted to derive the fundamental parameters for each author. SPh and 
JPh correspond to Str\"omgren and Johnson photometry respectively. "Fe eq" 
indicates the use of the Fe ionisation equilibrium, while H-fit the use of 
hydrogen line fitting. SED corresponds to the use of spectral energy 
distribution in the visible and/or UV region. All the values obtained from the 
literature are in excellent agreement with our adopted parameters. We would 
like to mention that only the oldest determination \citep{sadakane81} was 
obtained using all the possible parameter indicators, as done in this work.
\begin{table}[ht]
\caption[ ]{Atmospheric parameters of 21~Peg derived from other authors.}
\label{comp_param_21peg}
\begin{center}
\begin{tabular}{cccc}
\hline
\hline
Reference & \Teff & \logg & method \\
          &   [K] & [cgs] &        \\
\hline			        					       
S81   & 10500 & 3.50  & SPh, JPh, Fe eq, H-fit, SED \\
B82   & 10500 & 3.50  & SPh, H-fit \\
A83   & 10700 & (...) & JPh \\
A83   & 10350 & (...) & SPh \\
A83   & 10375 & (...) & SED \\
M85   & 10300 & (...) & SED \\
S93   & 10450 & 3.50  & SPh, H-fit \\
L98   & 10200 & 3.50  & SPh \\
W00   & 10350 & 3.48  & SPh \\
H01   & 10450 & 3.60  & SPh \\
A02   & 10375 & 3.47  & SPh \\
A02   & 10350 & 3.55  & SED, H-fit \\
F09   & 10400 & 3.55  & SED, H-fit \\
\hline
\end{tabular}
\end{center}
\smallskip 

S81: \citet{sadakane81}, B82: \citet{boesgaard82}, A83: \citet{adelman83},
M85: \citet{morossi85}, S93: \citet{smith93}, L98: \citet{john1998},
W00: \citet{wahlgren00}, H01: \citet{hubrig01}, A02: \citet{adelman02}
F09: this work. 
\end{table}


The spectroscopic literature for $\pi$~Cet is extremely vast and started in 
the early 60s. We decided to compare the determinations of \Teff\ and 
\logg\ from the late 70s on. These data and our determination are 
presented in Table~\ref{comp_param_picet}. Our value for \Teff\ is the only 
one below 13000\,K, while \logg\ agrees with all the other measurements. 
\citet{adelman02} also derived the parameters of $\pi$~Cet with the 
simultaneous fitting of H$\gamma$ and spectrophotometry, essentially the same 
way as we did. The main difference is that they only used one available 
spectrophotometric set, while we used all the data found in the literature. The 
spectrophotometry by \citet{breger} appears a little below the one by 
\citet{adelman1989} around the Balmer jump. To be able to simultaneously fit 
both sets of measurements, we needed a \Teff\ lower than 13000\,K and the best 
fit was obtained for \Teff\ = 12800\,K. Also our H$\beta$ profile is 
best-fitted only with a \Teff\ below 13000\,K, as already mentioned in 
Sect.~\ref{hlines}. 
\begin{table}[ht]
\caption[ ]{Atmospheric parameters of $\pi$~Cet derived from other authors.}
\label{comp_param_picet}
\begin{center}
\begin{tabular}{cccc}
\hline
\hline
Reference & \Teff & \logg & method \\
          &   [K] & [cgs] & (as in Table~\ref{comp_param_21peg})       \\
\hline			        					       
H79   & 13100 & 3.90  & SPh \\
K80   & 13000 & (...) & SED \\
M83   & 13030 & 3.88  & SED(UV) \\
M83   & 13170 & 3.91  & SED(visible) \\
A84   & 13150 & 3.65  & SED, H-fit \\
M85   & 13170 & (...) & SED \\
S88   & 13200 & 3.70  & SED \\
M88   & 13425 & (...) & SPh \\
R90   & 13150 & 3.50  & SPh \\
A91   & 13150 & 3.85  & H-fit \\
T91   & 13000 & (...) & SED \\
S92   & 13210 & 3.65  & SPh \\
A02   & 13174 & 3.70  & SPh \\
A02   & 13100 & 3.85  & SED, H-fit \\
F09   & 12800 & 3.75  & SED, H-fit \\
\hline
\end{tabular}
\end{center}
\smallskip

H79: \citet{heacox79}, K80: \citet{kontizas80}, M83: \citet{mala83},
A84: \citet{adelman84}, M85: \citet{morossi85}, S88: \citet{sadakane88}
M88: \citet{megessier88}, R90: \citet{roby90}, A91: \citet{adelman91}
T91: \citet{t91}, S92: \citet{s92}, A02: \citet{adelman02}
F09: this work. The methods  and the acronyms are as in
Table~\ref{comp_param_21peg}.
\end{table}

%
\section{Abundance analysis}\label{abundance analysis}
The \vald\ database is the main source for the atomic parameters of spectral 
lines. For light elements C, N, O, \ion{Ne}{i}, \ion{Mg}{i}, \ion{Si}{ii}, 
\ion{Si}{iii}, S, Ar, and also for \ion{Fe}{iii}, the oscillator strengths are 
taken from \nist\ online database \citep{NIST08}. LTE abundance analysis in 
the atmospheres of all three stars is based mainly on the equivalent widths 
analysed with the improved version of \width\ code updated to use the \vald\ 
output linelists and kindly provided to us by V.~Tsymbal. 

In the case of blended lines or when the line is situated in the 
wings of the hydrogen lines, we performed synthetic spectrum calculations 
with the \synth\ code. For our analysis we used the maximum number of 
spectral lines available in the observed wavelength ranges except 
lines in spectral regions where the continuum normalisation was too
uncertain (high orders of the Paschen series in the
\espa\ spectrum of $\pi$~Cet, for example). The final abundances are given in 
Table~\ref{abundance}. A line-by-line abundance list with the equivalent 
width measurements, adopted oscillator strengths, and their sources is given 
in Table~\ref{line_abund} (online material). Below we discuss the results 
obtained for individual elements.
\begin{table*}[ht]
\caption[ ]{LTE atmospheric abundances in programme stars with the error
estimates based on the internal scattering from the number of analysed lines,
$n$. }
\label{abundance}
\begin{center}
\begin{tabular}{l|cc|cc|cc|c}
\hline
\hline
Ion &\multicolumn{2}{|c|}{HD~145788} &\multicolumn{2}{c|}{21~Peg} &\multicolumn{2}{c|}{$\pi$~Cet} &  Sun (*) \\                                  
    &$\log (N/N_{\rm tot})$ & $n$  &$\log (N/N_{\rm tot})$ & $n$ &$\log (N/N_{\rm tot})$ & $n$ &$\log (N/N_{\rm tot})$  \\       
\hline
\ion{He}{i }  & ~~$-$1.10$\pm$0.05 &  4 & ~~$-$1.09$\pm$0.03 &  7 & ~~$-$0.97$\pm$0.04 &  6 & ~~$-$1.12~ \\                            
\ion{C}{i}    & ~~$-$3.60$\pm$0.14 &  5 & ~~$-$3.66$\pm$0.14 &  9 &                    &    & ~~$-$3.65~ \\			    
\ion{C}{ii}   & ~~$-$3.63:         &  2 & ~~$-$3.65$\pm$0.05 &  4 & ~~$-$3.58$\pm$0.07 &  7 & ~~$-$3.65~ \\			    
\ion{N}{i}    &                    &    & ~~$-$3.95$\pm$0.12 &  4 & ~~$-$4.03$\pm$0.13 & 10 & ~~$-$4.26~ \\			    
\ion{N}{ii}   &                    &    & ~~$-$3.90:         &  1 & ~~$-$3.74$\pm$0.07 &  9 & ~~$-$4.26~ \\			    
\ion{O}{i}    & ~~$-$3.12$\pm$0.09 &  7 & ~~$-$3.28$\pm$0.11 & 18 & ~~$-$3.06$\pm$0.14 &  9 & ~~$-$3.38~ \\			    
\ion{O}{ii}   &                    &    &                    &    & ~~$-$3.04:         &  2 & ~~$-$3.38~ \\			    
\ion{Ne}{i}   &                    &    & ~~$-$3.76$\pm$0.05 &  7 & ~~$-$3.66$\pm$0.09 & 20 & ~~$-$4.20~ \\			    
\ion{Na}{i}   &                    &    & ~~$-$5.60:         &  1 & ~~$-$5.23$\pm$0.07 &  3 & ~~$-$5.87~ \\			    
\ion{Mg}{i}   & ~~$-$4.17$\pm$0.24 &  5 & ~~$-$4.42$\pm$0.12 &  5 & ~~$-$4.27:         &  2 & ~~$-$4.51~ \\			    
\ion{Mg}{ii}  & ~~$-$4.29$\pm$0.06 &  4 & ~~$-$4.56$\pm$0.03 &  7 & ~~$-$4.47$\pm$0.16 & 10 & ~~$-$4.51~ \\			    
\ion{Al}{i }  & ~~$-$5.70:         &  2 & ~~$-$5.89:         &  2 & ~~$-$5.57:         &  2 & ~~$-$5.67~ \\			    
\ion{Al}{ii}  & ~~$-$5.11$\pm$0.10 &  3 & ~~$-$5.70$\pm$0.10 &  4 & ~~$-$5.73$\pm$0.27 &  8 & ~~$-$5.67~ \\			    
\ion{Al}{iii} &                    &    &                    &    & ~~$-$5.30$\pm$0.02 &  3 & ~~$-$5.67~ \\			    
\ion{Si}{i}   & ~~$-$4.75:         &  1 & ~~$-$4.95:         &  1 & ~~$-$4.80:         &  1 & ~~$-$4.53~ \\			    
\ion{Si}{ii}  & ~~$-$4.27$\pm$0.14 & 11 & ~~$-$4.49$\pm$0.13 & 22 & ~~$-$4.41$\pm$0.20 & 31 & ~~$-$4.53~ \\			    
\ion{Si}{iii} &                    &    & ~~$-$4.26$\pm$0.18 &  2 & ~~$-$4.16:         &  2 & ~~$-$4.53~ \\			    
\ion{P}{ii}   &                    &    & ~~$-$6.37$\pm$0.06 &  3 & ~~$-$6.38$\pm$0.19 &  9 & ~~$-$6.68~ \\			    
\ion{P}{iii}  &                    &    &                    &    & ~~$-$6.19:         &  1 & ~~$-$6.68~ \\			    
\ion{S}{ii}   & ~~$-$4.36:         &  2 & ~~$-$4.86$\pm$0.13 & 26 & ~~$-$4.78$\pm$0.16 & 31 & ~~$-$4.90~ \\			    
\ion{Cl}{ii}  &                    &    &                    &    & ~~$-$6.95:         &  2 & ~~$-$6.54~ \\			    
\ion{Ar}{i}   &                    &    &                    &    & ~~$-$4.86$\pm$0.24:&  2 & ~~$-$5.86~ \\			    
\ion{Ar}{ii}  &                    &    &                    &    & ~~$-$5.24$\pm$0.19 &  6 & ~~$-$5.86~ \\			    
\ion{Ca}{i}   & ~~$-$5.46$\pm$0.11 &  5 & ~~$-$5.84$\pm$0.11 &  3 &                    &    & ~~$-$5.73~ \\			    
\ion{Ca}{ii}  & ~~$-$5.54$\pm$0.16 &  6 & ~~$-$5.98$\pm$0.08 &  5 & ~~$-$5.77:         &  2 & ~~$-$5.73~ \\			    
\ion{Sc}{ii}  & ~~$-$8.90$\pm$0.03 &  5 & ~~$-$9.37$\pm$0.10 &  7 & ~~$-$9.31:         &  1 & ~~$-$8.99~ \\			    
\ion{Ti}{ii}  & ~~$-$6.80$\pm$0.15 & 47 & ~~$-$7.23$\pm$0.09 & 59 & ~~$-$7.42$\pm$0.08 & 11 & ~~$-$7.14~ \\			    
\ion{V}{ii}   & ~~$-$7.55$\pm$0.20 &  4 & ~~$-$7.98$\pm$0.06 &  9 &                    &    & ~~$-$8.04~ \\			    
\ion{Cr}{i}   & ~~$-$6.15$\pm$0.09 &  4 & ~~$-$6.29$\pm$0.09 &  5 &                    &    & ~~$-$6.40~ \\			    
\ion{Cr}{ii}  & ~~$-$5.86$\pm$0.11 & 31 & ~~$-$6.20$\pm$0.10 & 68 & ~~$-$6.41$\pm$0.10 & 21 & ~~$-$6.40~ \\			    
\ion{Mn}{i}   & ~~$-$6.43:         &  2 & ~~$-$6.54$\pm$0.21 &  5 &                    &    & ~~$-$6.65~ \\			    
\ion{Mn}{ii}  & ~~$-$6.15$\pm$0.08 &  4 & ~~$-$6.51$\pm$0.17 & 19 & ~~$-$6.50$\pm$0.09 &  3 & ~~$-$6.65~ \\			    
\ion{Fe}{i}   & ~~$-$4.23$\pm$0.16 & 88 & ~~$-$4.52$\pm$0.13 &108 & ~~$-$4.53$\pm$0.22 &  7 & ~~$-$4.59~ \\			    
\ion{Fe}{ii}  & ~~$-$4.13$\pm$0.15 &147 & ~~$-$4.50$\pm$0.12 &406 & ~~$-$4.58$\pm$0.14 &186 & ~~$-$4.59~ \\			    
\ion{Fe}{iii} &                    &    & ~~$-$4.60$\pm$0.06 &  3 & ~~$-$4.52$\pm$0.10 &  4 & ~~$-$4.59~ \\			    
\ion{Co}{ii}  &                    &    & ~~$-$6.75$\pm$0.18 &  3 & ~~$-$6.93:         &  1 & ~~$-$7.12~ \\			    
\ion{Ni}{i}   & ~~$-$5.32$\pm$0.13 &  8 & ~~$-$5.71$\pm$0.05 & 10 &                    &    & ~~$-$5.81~ \\			    
\ion{Ni}{ii}  & ~~$-$5.12$\pm$0.05 &  3 & ~~$-$5.61$\pm$0.09 & 23 & ~~$-$5.76$\pm$0.19 & 17 & ~~$-$5.81~ \\			    
\ion{Zn}{i}   &                    &    & ~~$-$6.86:         &  1 &                    &    & ~~$-$7.44~ \\			    
\ion{Sr}{ii}  & ~~$-$8.45:         &  2 & ~~$-$9.10:         &  2 & ~~$-$9.15:         &  2 & ~~$-$9.12~ \\			    
\ion{Y}{ii}   & ~~$-$9.06:         &  1 & ~~$-$9.76$\pm$0.15 &  4 &                    &    & ~~$-$9.83~ \\			    
\ion{Zr}{ii}  & ~~$-$8.75$\pm$0.32 &  3 & ~~$-$9.48$\pm$0.28 &  4 &                    &    & ~~$-$9.45~ \\			    
\ion{Ba}{ii}  & ~~$-$8.96$\pm$0.11 &  3 & ~~$-$9.19$\pm$0.06 &  3 &                    &    & ~~$-$9.87~ \\			    
\ion{Nd}{iii} &                    &    & ~$-$10.09$\pm$0.07 &  3 &                    &    & ~$-$10.59~ \\			    
\hline											     %
\Teff     &\multicolumn{2}{|c|}{9750~K}   &\multicolumn{2}{c|}{10400~K}   &\multicolumn{2}{c|}{12800~K}   & 5777~K  \\				
\logg     &\multicolumn{2}{|c|}{3.7~~~~~} &\multicolumn{2}{c|}{3.55~~~~}  &\multicolumn{2}{c|}{3.75~~~~} & 4.44~~~~\\				   
\hline											  
\end{tabular}
\end{center}
\smallskip

Internal scattering was not estimated when $n<3$, in which case
the derived abundance if flagged with a colon (:). (*) the abundances of the solar atmosphere
calculated by \citet{met05}.
\end{table*}

%
\subsection{Results for individual elements}
\subsubsection{Helium}\label{He}
Stark broadening of helium lines was treated using the \citet{BCS74} 
broadening theory and tables. For allowed isolated lines we used width and 
shift functions from Table 1 of this paper, while an interpolation of the 
calculated line profiles given in Tables 2-8 was employed for 
$\lambda$~4472\,\AA\ line. All abundances were derived without using 
equivalent widths, but by fitting of the line wings.

The quality of the fit to the observed \ion{He}{i} lines in the programme 
stars is demonstrated in Fig.~\ref{portions}, while the fit to 
\ion{He}{i}~$\lambda$~4472\,\AA\ line is shown in Fig.~\ref{He} for 
different temperatures in 21~Peg (Online material).

The helium abundance in HD~145788 and in 21~Peg is solar, while it is slightly 
overabundant in $\pi$~Cet. Our analysis was applied to the \ion{He}{i} lines 
at wavelengths shorter than 5000\,\AA, which should be influenced very little  
by \nlte\ effects, except, maybe, in the case of $\pi$~Cet where LTE 
synthetic profiles fit the line wings but not the line cores. Non-LTE 
calculations for \ion{He}{i} lines in the spectrum of $\beta$~Ori 
(\Teff=13000\,K, \logg=2.0) show that negative \nlte\ corrections of about 
0.1--0.2\,dex should even be applied to blue lines \citep{He1}. It is unclear 
what corrections are expected for main sequence stars of the same \Teff, 
therefore \nlte\ calculations for at least $\pi$~Cet are necessary for deriving 
He abundance with proper accuracy. Our high-quality observations of 
$\pi$~Cet will serve perfectly for a thorough \nlte\ study of \ion{He}{i} 
lines in middle B-type stars.
\subsubsection{CNO}
The carbon abundance as derived from \ion{C}{ii} lines is solar for all three 
stars. It is also very close to the cosmic abundance standard recently 
determined by \citet{P08} from the analysis of nearby early B-type stars and 
discussed in their paper. The ionisation equilibrium between \ion{C}{i} and 
\ion{C}{ii} deserves some short comment.

\citet{Prz-C} calculated \nlte\ corrections for \ion{C}{i} and
\ion{C}{ii} for Vega model atmosphere (\Teff=9550\,K, \logg=3.95). They
found these corrections to be negligible. \citet{rentzsch1996}
made a \nlte\ analysis of \ion{C}{ii} in A-type stars also obtaining very small
(less than 0.05\,dex) negative corrections at effective temperatures around 
10000\,K. Because the ionisation equilibrium between \ion{C}{i}/\ion{C}{ii} is 
fulfilled for HD 145788, we assume that \nlte\ corrections are also negligible 
for this star.

Instead, at higher \Teff\ values, the abundances obtained from \ion{C}{i} 
lines having the lower level 3s$^1$P$^{\rm o}$ 
($\lambda\lambda$~4932, 5052, 5380, 8335, 9406\,\AA) are significantly lower 
than those obtained from the other \ion{C}{i} lines. In the case of 21~Peg,
abundances of \ion{C}{i} and \ion{C}{ii} only agree if
we neglect the abundances obtained from $\lambda$~4932, 5052, 5380
lines. In the spectrum of the hottest star of our sample, $\pi$~Cet, the
situation is even more extreme. $\lambda\lambda$~4932, 5052, 5380
\ion{C}{i} lines are not visible at all, while $\lambda\lambda$~8335,
9406\,\AA\ lines appear in emission. The \ion{C}{i} lines at
$\lambda$~7111-7120\,\AA\ are rather shallow, we can only
determine an upper limit for the abundance: $\log (C/N_{\rm tot})$ =
$-4.0$. Like \citet{roby90} we also obtain a \ion{C}{i}/\ion{C}{ii} imbalance 
in $\pi$~Cet. Non-LTE calculations by \citet{rentzsch1996} seem to 
explain the unusual behaviour of \ion{C}{i} $\lambda\lambda$~4932, 5052, 5380 
lines in stars hotter than $\pi$~Cet because the abundance corrections become 
positive and grow with the effective temperature.

Nitrogen abundance is obtained for the two hottest stars of our programme 
from the lines of the neutral and singly-ionised nitrogen. While in 21~Peg we 
get the evidence for ionisation equilibrium, in $\pi$~Cet \ion{N}{ii} lines 
give higher abundance by 0.3\,dex. In both stars, LTE nitrogen abundance 
exceeds the solar one. For $\pi$~Cet, \citet{roby90} derived an approximately 
solar nitrogen abundance from \ion{N}{i} lines located between two Paschen 
lines. We use these lines, too, and the higher nitrogen abundance derived 
by us is caused by the larger equivalent widths, and not by the difference 
in the adopted effective temperature. From the \nlte\ calculations performed by 
\citep{Prz-N}, we may expect $-0.3$ abundance corrections for the lines of 
\ion{N}{i} that bring nitrogen abundance in both stars to the solar value. 
It is difficult to estimate corrections for \ion{N}{ii} lines. Evidently, 
\nlte\ analysis of both \ion{C}{i}/\ion{C}{ii} and \ion{N}{i}/\ion{N}{ii} 
line formation is necessary. 

Oxygen abundance was derived from the lines of neutral oxygen in HD~145788 
and in 21~Peg, while the lines of neutral and singly-ionised oxygen were 
used in $\pi$~Cet. Although there are plenty of \ion{O}{i} lines in the red 
region, our analysis was limited by the lines with $\lambda\le$~6500\,\AA, 
which are not influenced by \nlte\ effects or very little so \citep{Prz-O}. 
Even for $\lambda$~6155--8\,\AA, lines the abundance corrections are 
less than 0.1\,dex in main sequence stars. Within the errors of our abundance 
analysis, 21~Peg has nearly solar oxygen abundance. Moreover, it agrees 
perfectly with the cosmic abundance standard derived by \citet{P08}. In 
HD~145788 and $\pi$~Cet, oxygen seems to be slightly overabundant with values
falling in the solar photospheric range defined by \citet{grevesse1996} and 
\citet{met05}. For $\pi$~Cet our oxygen abundance agrees perfectly 
with that obtained by \citet{roby90}.
\subsubsection{Neon and argon}\label{NeAr}
The abundance of these noble gases in stellar atmospheres attracts special 
attention because they cannot be obtained directly in the solar atmosphere. 
These gases are volatile, and meteoritic studies also cannot provide the 
actual abundance in the solar system. The revision of the solar abundances by 
\citet{met05} that results in a 0.2--0.3\,dex decrease in CNO abundances, so 
those of other elements produced significant inconsistency between 
the predictions of the solar model and the helioseismology measurements. 
One of the ways to bring both data into agreement is to increase Ne abundance. 
Solar model calculations by \citet{bahcall05} show that $A$(Ne)=8.29$\pm$0.05 
is enough for this purpose. 

\citet{Cunha-Ne} performed \nlte\ analysis of neon line formation in the 
young B-type stars of the Orion association and derived an average 
$A$({\rm Ne})=8.11$\pm$0.05 ($\log ({\rm Ne}/N_{\rm tot})=-3.93$) from 11 
stars. \citet{HH2003} derived the Ne abundance in a sample of optically 
bright, early B-type main sequence stars, obtaining an average \nlte\ Ne 
abundance of $\log ({\rm Ne}/N_{\rm tot})=-3.93\pm0.13$. \citet{P08} 
obtained \nlte\ Ne abundances in a sample of six nearby main sequence early 
B-type stars. They derived an Ne abundance of $A$({\rm Ne})=8.08$\pm$0.03 
($\log ({\rm Ne}/N_{\rm tot})=-3.96$). All these values agree very 
well with $A$(Ne)=8.08$\pm$0.10 obtained from analysing the emission 
registered during low-altitude impulsive flare \citep{feldman90}, but are 
0.3\,dex higher than adopted by \citet{met05}. Finally, \ion{Ne}{i} and 
\ion{Ne}{ii} \nlte\ analysis in 18 nearby early B-type stars \citep{morel08} 
results in $A$(Ne)=7.97$\pm$0.07. All determinations in B-type stars agree 
within the quoted errors and provide the reliable estimate of neon abundance 
in the local interstellar medium, which is higher than the newly proposed solar 
neon abundance. 

Recently, \citet{Lanz-Ar} have studied the Ar abundance in the same set of 
young B-type stars of the Orion association and derived an average 
$A$({\rm Ar})=6.66$\pm$0.06 ($\log ({\rm Ar}/N_{\rm tot})=-5.38$), which again 
agrees with the value $A$(Ar)=6.57$\pm$0.12 obtained by \citet{feldman90}, 
but is $\sim$0.4\,dex higher than recommended by \citet{met05}. While \nlte\ 
effects on \ion{Ne}{i} lines are known to be strong 
\citep[see ][]{sigut99},
those on \ion{Ar}{ii} lines are weak, $\approx$0.03\,dex \citep{Lanz-Ar}.

We measured \ion{Ne}{i} lines in the spectra of 21~Peg and $\pi$~Cet and 
\ion{Ar}{i}/\ion{Ar}{ii} in $\pi$~Cet only. As expected, averaged LTE neon 
abundances in both stars are higher than the solar one and than that derived 
by \citet{HH2003}, \citet{Cunha-Ne}, \citet{morel08}, or \citet{P08} for 
B-type stars. However, applying \nlte\ corrections, calculated by 
\citet{Dworetsky00} for the strongest \ion{Ne}{i}~6402\,\AA\ line to 
our LTE abundances derived from this line in both stars we get 
$\log (Ne/N_{\rm tot})=-3.89$ and $-3.86$ for 21~Peg and $\pi$~Cet, 
respectively, which brings Ne abundance in both stars into rather good 
agreement with the results obtained for early B-type stars. Rough estimates of 
possible \nlte\ corrections in $\pi$~Cet for \ion{Ne}{i}~6402 and 6506\,\AA\ 
lines, for which LTE and \nlte\ equivalent widths versus effective temperature 
are plotted by \citet{sigut99}, give us $Ne_{LTE}-Ne_{NLTE}\sim-0.3$\,dex 
for each line, and it agrees with the correction $-$0.36 calculated by 
\citet{Dworetsky00} for \ion{Ne}{i}~6402 line. The star $\pi$~Cet is a young 
star and thus adds reliable current data on the Ne abundance, taking  
a large number of high-quality line profiles and secure model atmosphere 
parameters into account.

We derived argon abundance $\log ({\rm Ar}/N_{\rm tot})=-5.24\pm0.19$ in 
$\pi$~Cet from 5 weak but accurately measured \ion{Ar}{ii} lines. Within 
error bars this value agrees with the results by \citet{Lanz-Ar} for B-type 
stars in the Orion association. We also managed to measure the two strongest 
\ion{Ar}{i} lines at 8103 and 8115\,\AA. They each give an Ar abundance that 
is too high. The 8115\,\AA\ line is blended with a \ion{Mg}{ii} line, and 
taking this blend into account we get $\log ({\rm Ar}/N_{\rm tot})=-5.2$, 
while the 8103\,\AA\ line is too strong for its transition probability. 
Moreover, both lines are located in the region contaminated by weak 
telluric lines, therefore the extracted abundances may be uncertain. 
\subsubsection{Na, Mg, Al, Si, P, S, Cl}
In both 21~Peg and $\pi$~Cet, Na abundances are derived from lines
affected by \nlte\citep{takedaNa}, therefore it is not surprising that 
their measured values are different from the solar ones.

In all three stars we obtained a slight \ion{Mg}{i}/\ion{Mg}{ii}
imbalance. For 21~Peg and $\pi$~Cet, the abundance derived from
\ion{Mg}{ii} lines is close to solar. The abundance derived
from the weaker \ion{Mg}{i} lines is consistent to the one obtained from
\ion{Mg}{ii}, hence solar. We conclude that, for these two stars, 
Mg abundance is consistent with the solar one and that discrepancies 
observed in the strong \ion{Mg}{i} lines come from \nlte\ effects.

Magnesium is slightly overabundant in HD~145788, but all the \ion{Mg}{i} lines
are affected by \nlte\ effects. The sign and the magnitude of its effect
depends on both \Teff\ and \logg\ \citep{Prz-Mg}, which prevents us
from obtaining any firm conclusion, until detailed \nlte\ calculations
for \ion{Mg}{i} for early A- and middle B-type main sequence stars
are carried out.

Aluminum is above solar in HD~145788 and have nearly solar abundances
in the two other stars as derived from \ion{Al}{ii} lines. \ion{Al}{i}
lines in all stars and \ion{Al}{iii} lines in $\pi$~Cet provide some
discordant results. It is not possible to discuss the ionisation
equilibrium without a \nlte\ analysis of the line formation of all three ions.

Accurate silicon abundance determination in stars and the interstellar medium 
is an important part of abundance studies and intercomparisons because Si 
is a reference element for meteoritic abundances. Silicon is slightly above 
solar in HD~145788, and close to solar in 21~Peg and in 
$\pi$~Cet, if we consider the results obtained from the numerous \ion{Si}{ii} 
lines. The spectral synthesis in the region of the only 
\ion{Si}{i}~3905\,\AA\ line observed in all three stars provides much
lower Si abundance, and the abundance difference between \ion{Si}{i} and 
\ion{Si}{ii} is practically independent of \Teff. A \nlte\ analysis of Si 
line formation in the Sun and in Vega \citep{Wed01} shows that positive \nlte\ 
corrections are expected for \ion{Si}{i}~3905\,\AA\ line, while small negative 
corrections may be expected for \ion{Si}{ii} lines, thus leading both ions 
into the equilibrium. Our Si analysis is based on the very accurate 
transition probability for \ion{Si}{i}~3905\,\AA\ line \citep{BL91} and on 
a combination of transition probabilities extracted from a recent NIST 
compilation \citep{KP08} and theoretical calculations by \citet{AJPP81} 
for \ion{Si}{ii} lines. The NIST compilation does not contain data 
for about a quarter of the lines observed in our stars for which rather 
concordant data exist. Table~\ref{Si2-gf} (online material) gives a 
collection of the experimental, as well as theoretical, atomic parameters 
for \ion{Si}{ii} lines that may be useful in a future \nlte\ analysis. The 
dispersion in the measurements is of the order of the cited accuracies, and 
theoretical calculations agree rather well with the experimentally measured 
transition probabilities and Stark widths.        

Phosphorus is overabundant relative to the solar abundance by 0.3\,dex in
both hotter stars, 21~Peg and in $\pi$~Cet. Oscillator strengths for
\ion{P}{ii} lines (taken from \vald) originally come from
calculations by \citet{H}. Therefore, at least part of the
observed overabundance may be caused by uncertainties in calculated 
transition probabilities. No \nlte\ analysis is available for the
phosphorus line formation. At the limit of detection, we managed to measure 
the two strongest \ion{Cl}{ii} lines (at 4794, 4810\,\AA) in $\pi$~Cet. The 
obtained upper limit on the chlorine abundance is 0.4\,dex lower than the 
recommended solar value \citep{met05}. Sulphur is overabundant by 0.5\,dex 
in HD~145788 and almost solar in the two other stars.
\subsubsection{Ca and Sc}\label{CaSc}
These two elements are of special interest in A-type star studies, as
their non-solar abundances indicate of a star's classification as
a metallic-line (Am) star. In hot Am stars with \Teff\ close to
HD~145788, both elements, in particular scandium, are
underabundant by 0.4--0.5\,dex, while other Fe-peak elements are
overabundant by $\sim$0.2--0.3\,dex 
\citep[see for instance o~Peg, which is a typical representative of the hot Am stars,][]{adelman88}.

Calcium and scandium are overabundant in HD~145788, but underabundant
in 21~Peg by 0.2\,dex and 0.4\,dex, respectively. Formal Ca-Sc
classification criteria of Am stars would make 21~Peg the hottest known 
Am star. However, classical Am stars are also characterised by
overabundances of $\sim 0.2-0.3$\,dex for other Fe-peak elements, and even
more remarkable overabundances of Sr, Y, and Zr \citep{praesepe1}. All
these peculiarities are not observed in 21~Peg, which therefore cannot
be classified as Am. The star $\pi$~Cet has solar Ca abundance and the same Sc 
deficiency as 21~Peg. 

At the \Teff\ of 21~Peg \nlte\ corrections for both \ion{Ca}{i} and 
\ion{Ca}{ii}, lines are expected to be positive (L. Mashonkina, private 
communication). In other words, the Ca abundance obtained from LTE 
calculations is, perhaps, underestimated, which may explain the observed 
discrepancies with respect to the solar case. Detailed \nlte\ analysis 
of the formation of Ca lines is required for accurate abundances.

Including hyperfine splitting (\hfs) does not change abundance results
because \hfs-effects are negligible for the investigated \ion{Sc}{ii}
lines (see Kurucz' \hfs\ calculations
\footnote{\tt http://cfaku5.cfa.harvard.edu/ATOMS}). Scandium deficiency 
requires more careful \nlte\ analysis, because it is observed not only in 
classical Am stars, but also in other stars, for example, in the A-type 
supergiant Deneb \citep{deneb}, which have solar abundances of the other 
Fe-peak elements. 
\subsubsection{Ti, V, Mn, Co, Ni}
Within the error limits, all these elements are almost solar in
21~Peg. The same is true for $\pi$~Cet, except for Ti, which is
slightly underabundant. The Ti abundance determinations are
based on the accurate laboratory transition probabilities \citep{PTP}
currently included in \vald. The \nlte\ corrections are expected to be 
positive \citep{deneb}, leading to abundance values closer to the solar one.

The situation is a bit different in the atmosphere of HD~145788, where
all these elements exhibit 0.2--0.4\,dex overabundance relative to
the solar photospheric abundances. (This would indicate that the star is Am, 
but since Ca and Sc are not underabundant, the star cannot be classified 
as Am). Still in HD~145788, for Mn, Ni, Cr, and Fe, the lines of the first ions
provide slightly higher abundance than the lines of the neutrals,
while no significant difference in ionisation equilibrium is observed
in the two other stars. The Mn lines are known to have rather large \hfs. We
checked the influence of \hfs\ on the derived Mn abundances for
HD~145788 where we measured the largest equivalent widths. Data
on \hfs\ for Mn were taken from \citet{Mn1hfs05} (\ion{Mn}{i}) and
from \citet{HSR99} (\ion{Mn}{ii}). We found that this effect is weak,
and does not exceed 0.05\,dex even for the lines with the largest \hfs.
\subsubsection{Cr}  
Although based only on a few lines, abundances derived from \ion{Cr}{i} lines 
are accurate because the recommended atomic parameters of these transitions 
\citep{MFW} currently included in \vald\ are supported by recent precise 
laboratory measurements \citep{sobeck07}.

For \ion{Cr}{ii}, laboratory measurements are only available for the
low-lying lines with $E_{\rm }i\le$4.8\,eV. Few measured lines in
HD~148788 and in $\pi$~Cet, and about one third of \ion{Cr}{ii} lines
in 21~Peg, originated in levels with higher excitation
potential. For these lines, only theoretical calculations are
available. In several publications, favour was given to the transition
probability calculations performed with the orthogonal operator
technique \citep[RU:][]{RU}, which are collected in the RU
database\footnote{ftp://ftp.wins.uva.nl/pub/orth} for the \ion{Cr}{ii},
\ion{Fe}{ii}, and \ion{Co}{ii} ions. All these data are included in the
current version of \vald. A comparison between RU calculations and the
most recent laboratory analysis of 119 lines of \ion{Cr}{ii}  
in 2055--4850\,\AA\ spectral region \citep{NLLN} shows that both sets
agree within 10\% on the absolute scale with a dispersion of 0.13\,dex. 
In the optical spectral region, \citet{NLLN} measures only 7
lines (in 4550--4850\,\AA\ region). To provide a consistent analysis,
we decided to use RU data for all lines of \ion{Cr}{ii} in our work.
Ionisation equilibrium was found for 21~Peg, while a slight imbalance
was found between \ion{Cr}{i} and \ion{Cr}{ii} in HD~145788. In
the spectrum of $\pi$~Cet no \ion{Cr}{i} lines are present.  

We found that Cr is overabundant in HD~145778 by 0.4\,dex (as average 
between the \ion{Cr}{i} and \ion{Cr}{ii} abundances), slightly overabundant 
in 21~Peg (by 0.15\,dex), while it has solar abundance in $\pi$~Cet.
\subsubsection{Fe}\label{Fe}
In 21~Peg and in $\pi$~Cet, iron abundance is practically solar. In
particular, the same Fe abundance is derived from Fe lines in three 
ionisation stage for $\pi$~Cet. No obvious ion imbalance was detected.

Iron is a crucial element for adjusting microturbulence, metallicity, and 
model atmosphere parameters. It has the most spectral
lines in the first three ionisation stages with relatively accurate
atomic data that can be observed in the optical spectra of early A- and
middle B-type stars. Numerous \ion{Fe}{ii} lines in the range of
excitation energy 2.5--11.3\,eV are seen in the spectrum of 21~Peg. In
the 4000--8000\,\AA\ spectral region, laboratory measurements are
available for 66 \ion{Fe}{ii} lines with $E_{\rm i}\le$6.2\,eV
\citep{vald3}. We measured 406 \ion{Fe}{ii} lines in the 21~Peg spectrum, 
and more than 200 lines have excitation potential
$>10$\,eV. In $\pi$~Cet, 98 out of 186 measured \ion{Fe}{ii} lines
have excitation potential $>10$\,eV. For all these high-excitation
lines, atomic parameters are available only through
theoretical calculations. As for \ion{Cr}{ii} lines, we checked the
RU database. For 66 lines for which laboratory transition
probabilities are measured, a comparison with the RU data results in
\loggf (lab data) $-$ \loggf (RU data) = $0.11 \pm 0.11$\,dex. This means that
using the \ion{Fe}{ii} lines of the homogeneous set of transition
probabilities obtained from theoretical calculations and available in
the RU database, we may over or underestimate the corresponding
abundances by no more than 0.1\,dex. Only calculated transition 
probabilities are available for \ion{Fe}{iii} lines. 

For a comparison of the accuracy of \ion{Fe}{ii} lines atomic
parameters, we obtained three sets of abundance determinations
for 21~Peg: one based on laboratory data included in \vald, one based on 
the recent NIST compilation \citep{NIST08}, and
one based on RU data. The results are (i) from laboratory \vald\ data: $\log
(\ion{Fe}{ii}/N_{\rm tot})= -4.61 \pm 0.11$ (51 lines); (ii) from NIST 
data: $\log (\ion{Fe}{ii}/N_{\rm tot})= -4.52 \pm 0.17$ (68 lines); (iii) from 
the same set of RU data: $\log (\ion{Fe}{ii}/N_{\rm tot})= -4.49 \pm 0.09$ 
(51 lines). Few strong high-excitation \ion{Fe}{ii} lines are included in the 
NIST compilation. These results justify the use of RU calculations for 
accurate iron abundance analysis, when laboratory measurements are not 
available. 

High accuracy of spectral data, together with very low \vsini, and fairly 
well-established atmospheric parameters, makes 21~Peg a
perfect object for an \ion{Fe}{ii} study. We could measure practically
all unclassified lines with intensity 1 and higher given in the
list of laboratory measurements \citep{johansson78}. These lines
belong to the transitions with very high excitation
potentials. Accurate position and intensity measurements in stellar
spectra may help in further studies of the \ion{Fe}{ii} spectrum and term
system.

Discussion of the possible \nlte\ effects on \ion{Fe}{i} and 
\ion{Fe}{ii} lines was given at the end of Sect.~\ref{metal_lines}. 
LTE iron abundances in 21~Peg and in $\pi$~Cet agree well with the cosmic 
abundance standard \citep[see][]{P08}.
\subsubsection{Sr, Y, Zr}
These elements are overabundant in HD~145788 and have solar abundances in 
21~Peg and $\pi$~Cet. For the Zr analysis we used the most recent 
experimental transition probabilities from \citet{LNAJ}.
\subsubsection{Ba and Nd}\label{BaNd}
Barium is overabundant in HD~145788 and in 21~Peg. No lines of elements heavier 
than zirconium are identified in our hottest programme star $\pi$~Cet. 

While barium overabundance, together with strontium-yttrium-zirconium 
overabundances in HD~145788, favours its classification as a hot Am star, 
barium overabundance in 21~Peg, which otherwise has near solar abundances of 
practically all other elements, is unexpected. Non-LTE corrections to barium 
abundance, if any, should be positive (L. Mashonkina, private communication). 
\citet{gigas88} derived for Vega \nlte\ corrections for the two barium lines 
($\lambda\lambda$~4554, 4934\,\AA) measured in this work as well. They 
obtained a positive correction of about 0.3\,dex for both lines. The \nlte\ 
corrections that should be applied to the barium abundance obtained in 
HD~145788 and in 21~Peg make the problem of the barium overabundance even 
more puzzling. Practically in all the abundance studies of normal A-type stars 
Ba was found to be overabundant \citep{lemke90,HL1993}. 

We measured three weak features at the position of the strongest \ion{Nd}{iii} 
lines \citep{RRKB} that result in slight Nd overabundance in 21~Peg. While Nd 
overabundance may still be attributed to the uncertainties on the absolute 
scale for calculated transition probabilities or \nlte\ effects, which are 
expected to be negative \citep{MRR05}, it is not the case for Ba where 
the atomic parameters of the lines using in our analysis are accurately known 
from laboratory studies.   
\subsection{Abundance uncertainties}
The abundance uncertainties for each ion shown in Table~\ref{abundance} are 
the standard deviation of the mean abundance obtained from the individual 
line abundances. 
Since our derivation of the abundances is mainly based on equivalent widths, 
we first have to estimate equivalent width errors given a certain SNR and 
\vsini. With a two $\sigma$ error bar, we derived 1.2\,m\AA\ for HD~145788, 
0.2\,m\AA\ for 21~Peg and 0.5\,m\AA\ for $\pi$~Cet. These values were derived 
by assuming a triangular line with a depth (height of the triangulum) equal to 
2$\sigma$ (SNR) and a width (base of the triangulum) equal to $2\times$\vsini. 
The rather high uncertainty on the equivalent widths of HD~145788 is mainly 
due to the low SNR of its spectrum. This shows the importance of a very high 
SNR not only for fast rotating stars, but also for slowly rotating stars. This 
uncertainty includes the uncertainty due to the continuum normalisation.  

\begin{figure}[ht]
\begin{center}
\includegraphics[width=90mm]{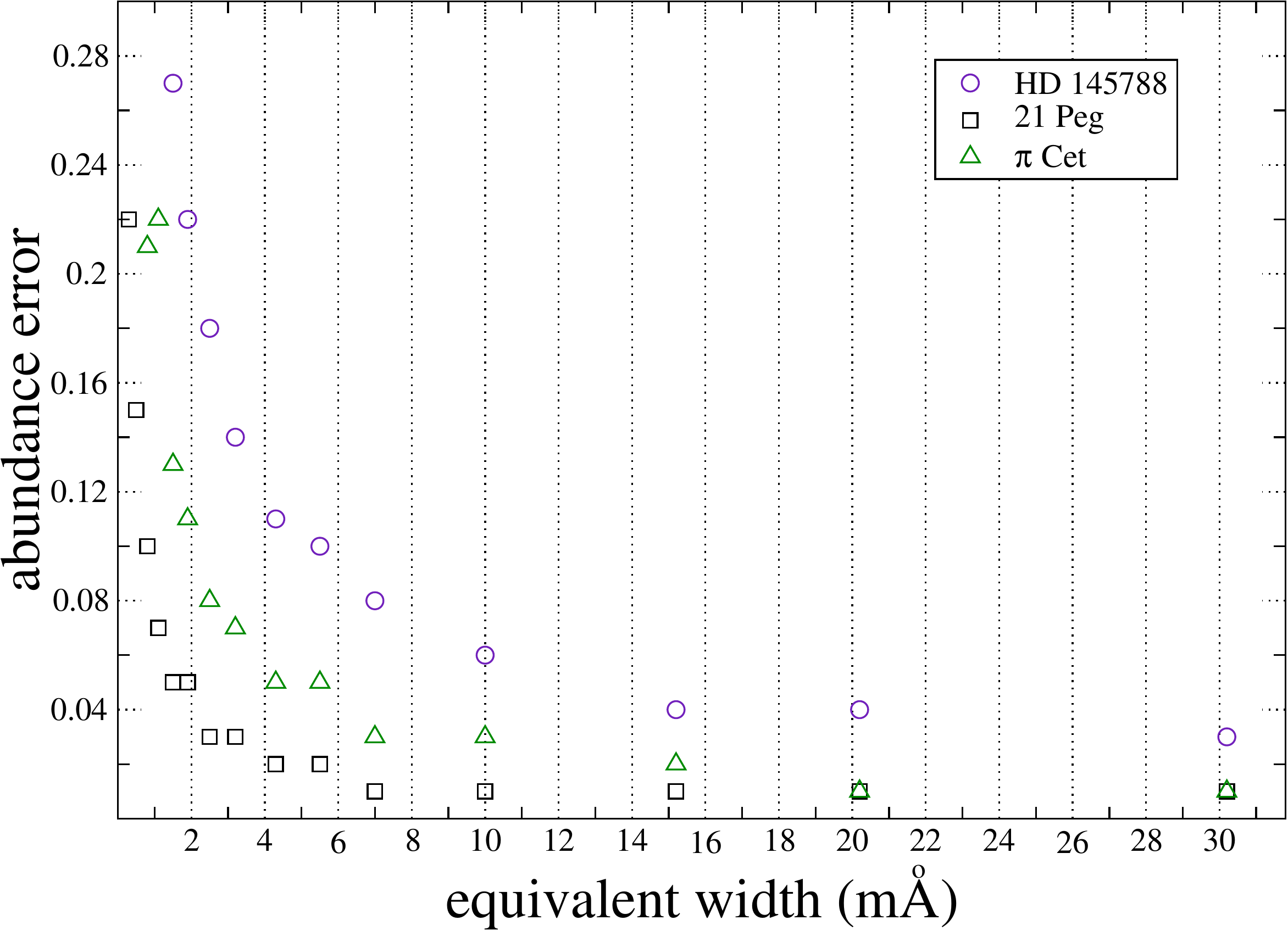}
\caption{Error bar in abundance as a function of equivalent widths for HD~145788
(open circle), 21~Peg (open square), and $\pi$~Cet (open triangle). This 
uncertainty stems from the uncertainty in the equivalent-width measurement and
continuum normalisation. The difference in uncertainty between 21~Peg and
$\pi$~Cet comes almost only from the difference in \vsini, while the high
error obtained for HD~145788 mainly from the low SNR of our observations,
relative to one of the other two stars.} 
\label{abn_error_eqw} 
\end{center} 
\end{figure}

In Fig.~\ref{abn_error_eqw} we plotted the error bars in abundance for a 
given line as a function of equivalent widths, for all the three stars 
analysed in this work. To derive the uncertainty in abundance due to the 
error bar on equivalent widths measurement, we took a representative 
\ion{Fe}{ii} line and derived the abundance of this line on the basis of 
different values of equivalent widths ranging from 0.3\,m\AA\ to 110\,m\AA. 
We then calculated the difference between the abundance obtained with the 
equivalent width X and X+$\delta$X, where $\delta$X is the error bar on the 
equivalent widths measurement. As expected, HD~145788 is the star that shows 
the largest error bar. For equivalent widths greater than 30\,m\AA, the 
abundance error tends asymptotically to zero. Table~\ref{line_abund} also 
shows that all the lines measured for this work are above the detection limit 
given by the \vsini\ and the SNR.

The mean equivalent width measured in these three stars is about 20\,m\AA\ for
HD~145788, 5\,m\AA\ for 21~Peg, and 8\,m\AA\ for $\pi$~Cet. These values 
correspond to an error bar in abundance, because of the uncertainty on the 
equivalent widths measurement and continuum normalisation of 0.04\,dex for 
HD~145788, and 0.03\,dex for both 21~Peg and $\pi$~Cet.

When we have enough measured lines, we assume that the 
internal scatter for each ion also takes the errors due to 
equivalent widths measurement and continuum normalisation into account.

\begin{figure}[ht]
\begin{center}
\includegraphics[width=90mm]{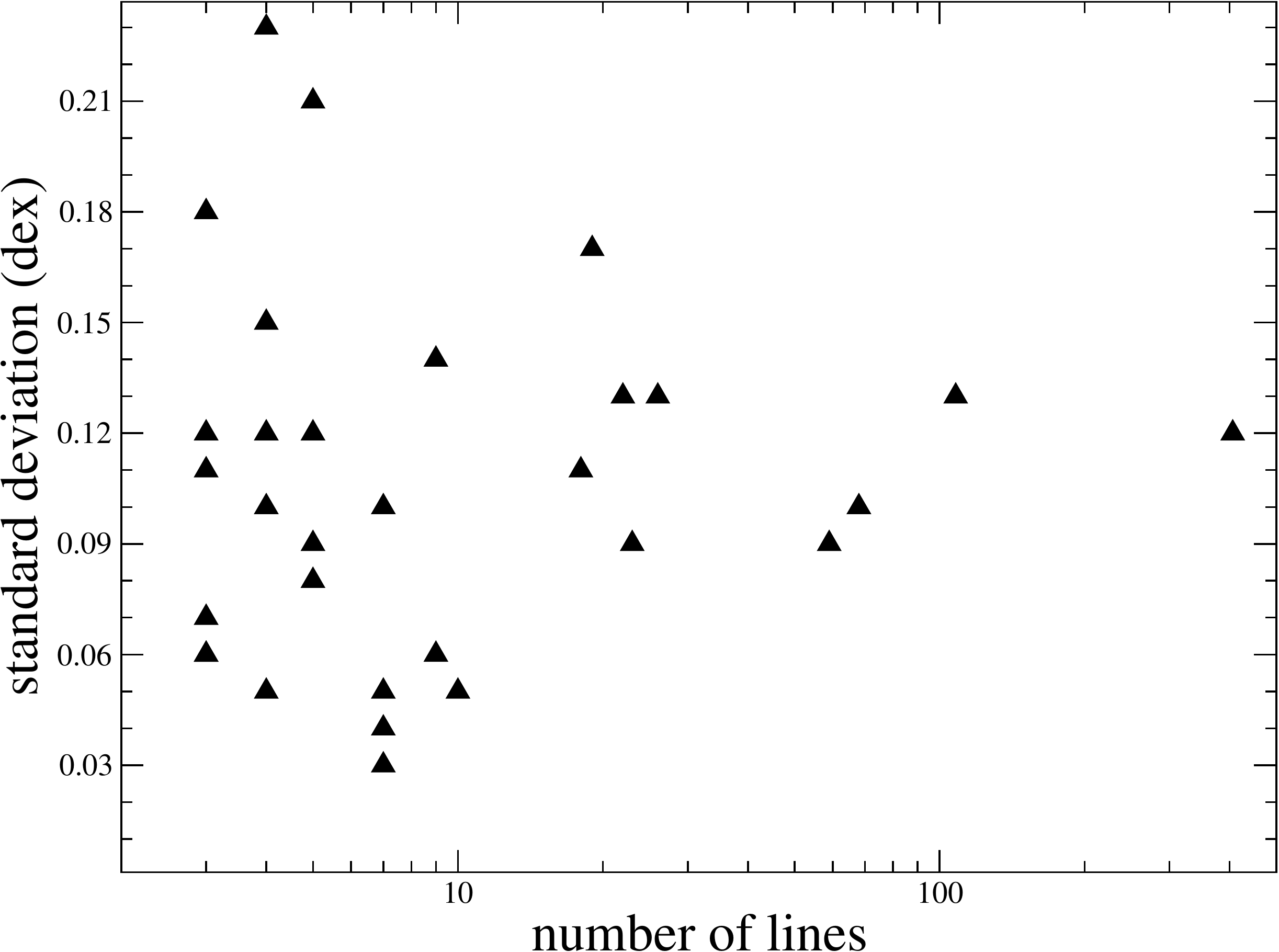}
\caption{Standard deviation of the derived abundances as a function of the 
the number of lines (shown in logarithmic scale and for a number of lines 
greater than 2). For visualisation reasons we omitted the standard deviation 
given by \ion{Zr}{ii}.} 
\label{error_numberlines} 
\end{center} 
\end{figure}

Figure~\ref{error_numberlines} shows the abundance scatter as a function of 
the number of measured lines for 21~Peg. For elements where \nlte\ 
effects are supposed to be important and line-dependent, such as Al and
S, the internal standard deviation is particularly high. The same is
found for elements with lower accuracy in \loggf\ values
due to the complexity of the atomic levels, such as Si. For 
other elements with a large enough number of spectral lines, 
say $n>10$, it is reasonable to expect an internal error of 0.11\,dex 
(see Fig.~\ref{error_numberlines}).

Considering the errors in oscillator strengths determination 
(see Table~\ref{line_abund}) given for the laboratory data, we may say that 
these errors are smaller for most elements than the internal scatter, so 
do not significantly influence the final results. The same is relevant for 
theoretical calculations. As already shown for \ion{Cr}{ii} and 
\ion{Fe}{ii} (Sect.~\ref{Fe}), calculated and laboratory sets of oscillator 
strengths agree within 0.1\,dex.

It should be noted that error due to the internal scatter is just
a part of the total error bar on the abundance determination.
To derive a more realistic abundance uncertainty we also have
to take the error bar due to systematic uncertainties in the 
fundamental parameters into account.
For a detailed discussion of the abundance error bars we again use 21~Peg. 
Table~\ref{error} shows the variation in abundance for each analysed ion, 
caused by the change of one fundamental parameter by $+1\sigma$, keeping fixed 
the other parameters. 
For the comparison $\sigma_{\rm abn}^2$\,(syst.) with 
$\sigma_{\rm abn}$\,(scatt.) of those ions for which the internal scattering 
could not be measured, an \textit{a priori} scatter of 
$\sigma_{\rm abn}\,{\rm (scatt.)} = 0.11$\,dex has to be assumed.
\begin{table*}[ht]
\caption{
Error sources for the abundances of the chemical elements of 21~Peg.
}
\label{error}
\begin{center}
\begin{tabular}{lrrrrrr}
\hline
\hline
\multicolumn{1}{c}{Ion         }        &  
\multicolumn{1}{c}{abundance   }        &  
\multicolumn{1}{c}{$\sigma_{\rm abn}$\,(scatt.)}  &  
\multicolumn{1}{c}{$\sigma_{\rm abn}$\,(\Teff) }  &  
\multicolumn{1}{c}{$\sigma_{\rm abn}$\,(\logg) }  &  
\multicolumn{1}{c}{$\sigma_{\rm abn}$\,(\vmic) }  &  
\multicolumn{1}{c}{$\sigma_{\rm abn}$\,(syst.) }  \\ 
\multicolumn{1}{c}{            }        & 
\multicolumn{1}{c}{$\log (N/N_{\rm tot})$} & 
\multicolumn{1}{c}{(dex)       }        & 
\multicolumn{1}{c}{(dex)       }        & 
\multicolumn{1}{c}{(dex)       }        & 
\multicolumn{1}{c}{(dex)       }        & 
\multicolumn{1}{c}{(dex)       }        \\ 
\hline
\ion{He}{i }  & $-$1.11 & 0.04 & $-$0.08 &    0.03 &    0.00 & 0.09 \\		  
\ion{C}{i}    & $-$3.66 & 0.14 & $-$0.05 & $-$0.13 & $-$0.10 & 0.17 \\		  
\ion{C}{ii}   & $-$3.65 & 0.05 & $-$0.17 & $-$0.02 & $-$0.08 & 0.19 \\		  
\ion{N}{i}    & $-$3.95 & 0.12 & $-$0.05 & $-$0.07 & $-$0.05 & 0.10 \\		  
\ion{N}{ii}   & $-$3.90:&      & $-$0.09 &    0.05 &    0.00 & 0.10 \\		  
\ion{O}{i}    & $-$3.28 & 0.11 &    0.00 & $-$0.02 & $-$0.02 & 0.03 \\	      
\ion{Ne}{i}   & $-$3.76 & 0.05 & $-$0.18 & $-$0.03 & $-$0.06 & 0.19 \\	      
\ion{Na}{i}   & $-$5.60 &      &    0.04 & $-$0.04 & $-$0.01 & 0.06 \\	      
\ion{Mg}{i}   & $-$4.42 & 0.12 &    0.10 & $-$0.06 & $-$0.05 & 0.13 \\	      
\ion{Mg}{ii}  & $-$4.56 & 0.03 & $-$0.02 & $-$0.01 & $-$0.02 & 0.03 \\	      
\ion{Al}{i }  & $-$5.89 &      &    0.10 & $-$0.03 &    0.00 & 0.10 \\	      
\ion{Al}{ii}  & $-$5.68 & 0.10 & $-$0.09 &    0.01 & $-$0.03 & 0.10 \\	      
\ion{Si}{i}   & $-$4.95 &      &    0.10 & $-$0.02 &    0.00 & 0.10 \\	      
\ion{Si}{ii}  & $-$4.49 & 0.13 & $-$0.07 &    0.01 & $-$0.03 & 0.08 \\	      
\ion{Si}{iii} & $-$4.26 & 0.18 & $-$0.16 &    0.04 & $-$0.03 & 0.17 \\	      
\ion{P}{ii}   & $-$6.37 & 0.06 & $-$0.08 &    0.02 & $-$0.02 & 0.08 \\	      
\ion{S}{ii}   & $-$4.86 & 0.13 & $-$0.14 &    0.00 & $-$0.05 & 0.15 \\	      
\ion{Ca}{i}   & $-$5.84 & 0.11 &    0.20 & $-$0.07 &    0.00 & 0.21 \\	      
\ion{Ca}{ii}  & $-$5.98 & 0.08 &    0.03 & $-$0.05 & $-$0.04 & 0.07 \\	      
\ion{Sc}{ii}  & $-$9.37 & 0.10 &    0.12 & $-$0.01 &    0.00 & 0.12 \\	      
\ion{Ti}{ii}  & $-$7.23 & 0.09 &    0.08 &    0.00 & $-$0.02 & 0.08 \\	      
\ion{V}{ii}   & $-$7.98 & 0.06 &    0.06 &    0.02 &    0.00 & 0.06 \\	      
\ion{Cr}{i}   & $-$6.29 & 0.09 &    0.13 & $-$0.03 &    0.00 & 0.13 \\	      
\ion{Cr}{ii}  & $-$6.20 & 0.10 &    0.03 &    0.02 & $-$0.01 & 0.04 \\	      
\ion{Mn}{i}   & $-$6.54 & 0.21 &    0.14 & $-$0.03 &    0.00 & 0.14 \\	      
\ion{Mn}{ii}  & $-$6.51 & 0.17 &    0.01 &    0.01 & $-$0.01 & 0.01 \\	      
\ion{Fe}{i}   & $-$4.52 & 0.13 &    0.11 & $-$0.03 &    0.00 & 0.11 \\	      
\ion{Fe}{ii}  & $-$4.50 & 0.12 & $-$0.02 &    0.02 & $-$0.02 & 0.03 \\	      
\ion{Fe}{iii} & $-$4.60 & 0.06 & $-$0.10 &    0.06 & $-$0.01 & 0.12 \\	      
\ion{Co}{ii}  & $-$6.75 & 0.18 &    0.01 &    0.03 &    0.00 & 0.03 \\	      
\ion{Ni}{i}   & $-$5.71 & 0.05 &    0.09 & $-$0.03 & $-$0.01 & 0.10 \\	      
\ion{Ni}{ii}  & $-$5.61 & 0.09 & $-$0.04 &    0.03 & $-$0.01 & 0.05 \\	      
\ion{Zn}{i}   & $-$6.86 &      &    0.10 & $-$0.02 &    0.00 & 0.10 \\	      
\ion{Sr}{ii}  & $-$9.10 & 0.01 &    0.13 & $-$0.03 & $-$0.06 & 0.15 \\	      
\ion{Y}{ii}   & $-$9.76 & 0.15 &    0.13 & $-$0.02 & $-$0.01 & 0.13 \\	      
\ion{Zr}{ii}  & $-$9.48 & 0.28 &    0.11 &    0.00 &    0.00 & 0.11 \\	      
\ion{Ba}{ii}  & $-$9.19 & 0.06 &    0.14 & $-$0.02 & $-$0.01 & 0.14 \\	      
\ion{Nd}{iii} &$-$10.09 & 0.07 &    0.04 &    0.04 &    0.00 & 0.06 \\	      
\hline								
\end{tabular}
\end{center}
\smallskip

Column 3 standard deviation $\sigma_{\rm abn}$ (scatt.) 
of the mean abundance obtained
from different spectral lines (internal scattering); a blank means that the number
of spectral lines is $<3$, hence no internal scattering could be estimated.
(Note that these values are identical to those given in Table~\ref{abundance}.)
Columns 4, 5, and 6 give the variation in abundance estimated by increasing
\Teff\ by 200\,K, \logg\ by 0.1\,dex, and \vmic\ by 0.4\,km\,s$^{-1}$,
respectively. Column~7 gives the the mean error calculated applying the
standard error propagation theory on the systematic uncertainties given in 
columns 4, 5 and 6, i.e.,
$\sigma_{\rm abn}^2$\,(syst.)    = 
$\sigma_{\rm abn}^2$\,(\Teff)  + 
$\sigma_{\rm abn}^2$\,(\logg)  + 
$\sigma_{\rm abn}^2$\,(\vmic).
\end{table*}


The main source of uncertainty is the error in the effective temperature 
determination, while the variation due to \logg\ and, in particular, to \vmic\ 
is almost negligible, although here we are considering an uncertainty on 
\vmic\ of 2$\sigma$. The abundance variation due to an increase in \logg\ 
leads to a worse ionisation equilibrium for many elements for which the 
equilibrium is reached at the adopted \logg, such as Cr, Mn, Fe, and Ni.
We repeated the same analysis for \ion{Fe}{ii} for HD~145788 and $\pi$~Cet. 
The results are comparable with those obtained for 21~Peg. We also expect the 
same effect for the other ions.

Assuming the different errors in the abundance determination are independent, 
we derived the final error bar using standard error propagation theory, 
given in column six of Table~\ref{error}. When the abundance is given by a
single line, we assumed an internal error of 0.11\,dex. Using the propagation 
theory we considered the situation where the determination of each fundamental 
parameter is an independent process. The mean value of the LTE uncertainties 
given in column six of Table~\ref{error} is 0.16\,dex. 

Because of the high SNR of the observations, the low \vsini\ and the non 
peculiarity of the programme stars, 
we can reasonably believe 
that the errors in abundance determinations estimated in the present study
are the smallest ones that could be obtained with the current state of the 
art of spectral LTE analysis for early-type stars. In the cases of stars with 
\vsini\ higher than those considered in this work, the uncertainty on the 
abundances increases. A higher rotational velocity would force the abundance 
analysis to be based on strong and saturated lines that are more sensitive 
to \vmic\ variations than weak lines. For a more detailed discussion, see 
\citet{luca2}.
\subsection{Comparison with previous abundance determinations}\label{comp_abn}
Table~\ref{comp_abn_21peg} (online) collects all previous massive 
abundance determinations in the atmosphere of 21~Peg 
\citep{sadakane81,smith93,smith1993,smith1994,Dworetsky00} in comparison with 
the results of the current analysis. 
We do not include works that only give abundances for a Cr and/or Fe. 

The main advantage of our analysis over the previous ones is the
wide wavelength coverage and the high quality of our spectra. 
It allows us to use many more spectral lines including very weak ones of the 
species not analysed before. Our analysis provides homogeneous abundance data 
for 38 ions of 26 chemical elements from He to Nd.  

Reasonably good agreement exists between our abundances derived with the 
spectra in optical and IR spectral regions and those derived with the UV 
observations \citep{smith93,smith1993,smith1994}, supporting the 
correctness of the adopted model atmosphere. 

\citet{Dworetsky00} derived the LTE neon abundance and gave \nlte\ abundance 
corrections for the strongest \ion{Ne}{i}~$\lambda$~6402 line. If we apply 
the \nlte\ correction described in Sect.~\ref{NeAr} to this line, we almost 
get the same \nlte\ abundance.  

For $\pi$~Cet the number of abundance determinations in the literature is
particularly vast. As for 21~Peg we consider only those where abundances are
derived for large number of ions: \citet{adelman91}, 
\citet{smith93}, \citet{smith1993}, \citet{smith1994}, and \citet{acke04}, 
except for neon abundance where we again included LTE and \nlte\ results by 
\citet{Dworetsky00}. For the comparison (see the online 
Table~\ref{comp_abn_picet}) of the abundances obtained for $\pi$~Cet, we also 
show the adopted \vmic\ because we believe that differences in this parameter 
are responsible for the difference between our abundances and those of 
\citet{acke04}. Again, we emphasise that the total number of ions (36) and 
elements (22), as well as the number of lines per ion analysed in the present 
work, is much higher than in any previous study. For Al, Si, Fe, we managed 
to derive abundances from the lines of the element in three ionisation stages, 
which provides a unique possibility to study \nlte\ effects. 

For the ions having several 
lines (\ion{S}{ii},
\ion{Ti}{ii}, \ion{Cr}{ii} and \ion{Fe}{ii}) we generally get a good agreement 
among the different authors, in particular, for \ion{Fe}{ii}.
The lower abundances given by \citet{acke04} are due 
to the high \vmic\ adopted.

The difference in He abundance between our work and that by \citet{adelman91} 
may be explained by the difference in \Teff, already discussed in 
Sect.~\ref{comp_param}. Both LTE and \nlte\ neon abundances agree rather well 
with the results by \citet{Dworetsky00} after applying the \nlte\ corrections 
given by these authors.
\section{Discussion}\label{discussion}
\begin{figure*}[ht]
\sidecaption
\includegraphics[width=140mm]{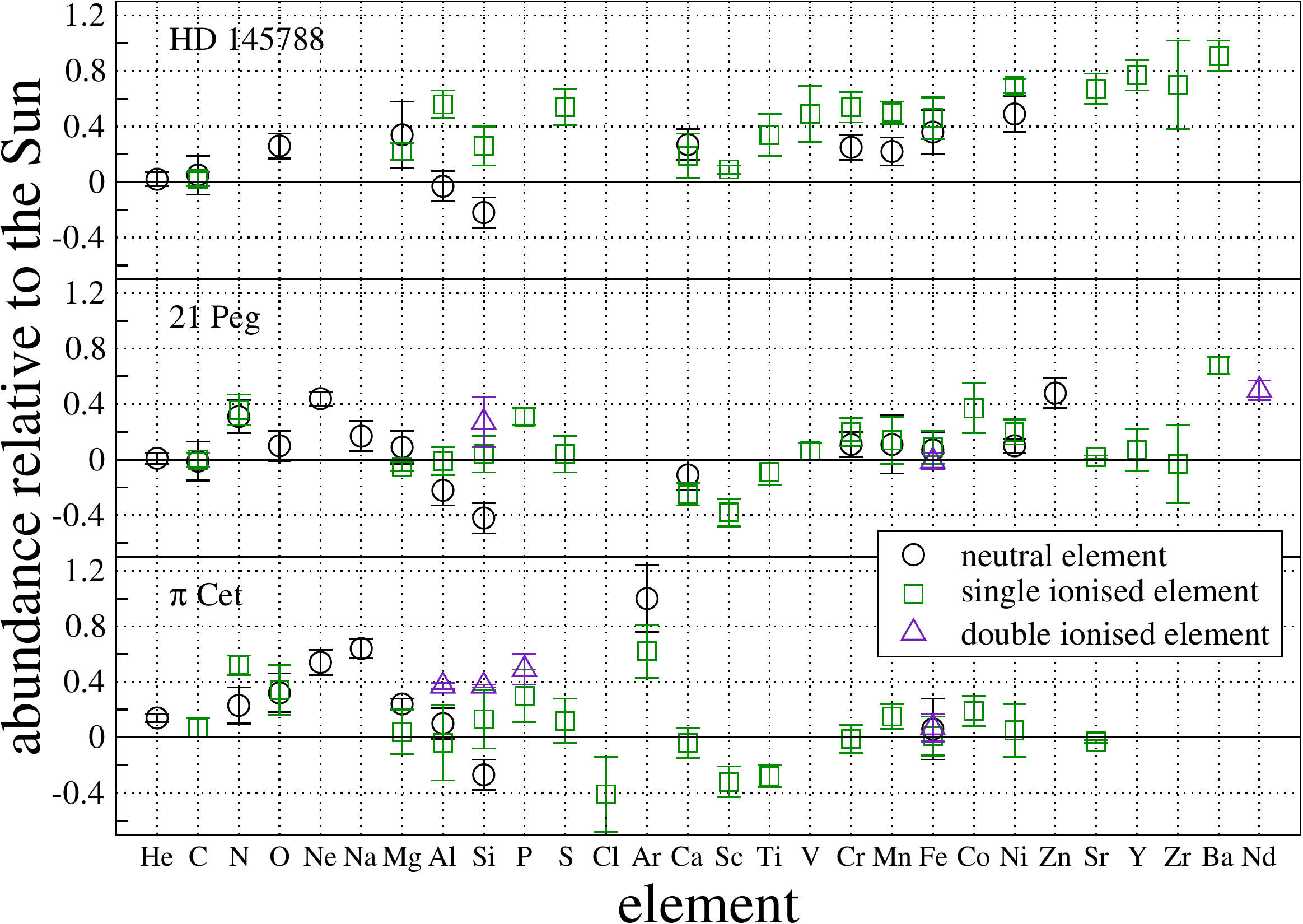}
\caption{LTE abundances relative to the Sun \citep{met05} for HD~145788, 
21~Peg, and $\pi$~Cet, from top to bottom. The open circles, squares, and 
triangles indicate the abundance for the neutral element, single ionised, and 
double ionised, respectively. The possible \nlte\ corrections for each ion are
described in the text from Secs.~\ref{He} to \ref{BaNd}.} 
\label{alltosun} 
\end{figure*}

\begin{figure*}[ht]
\begin{center}
\includegraphics[width=145mm]{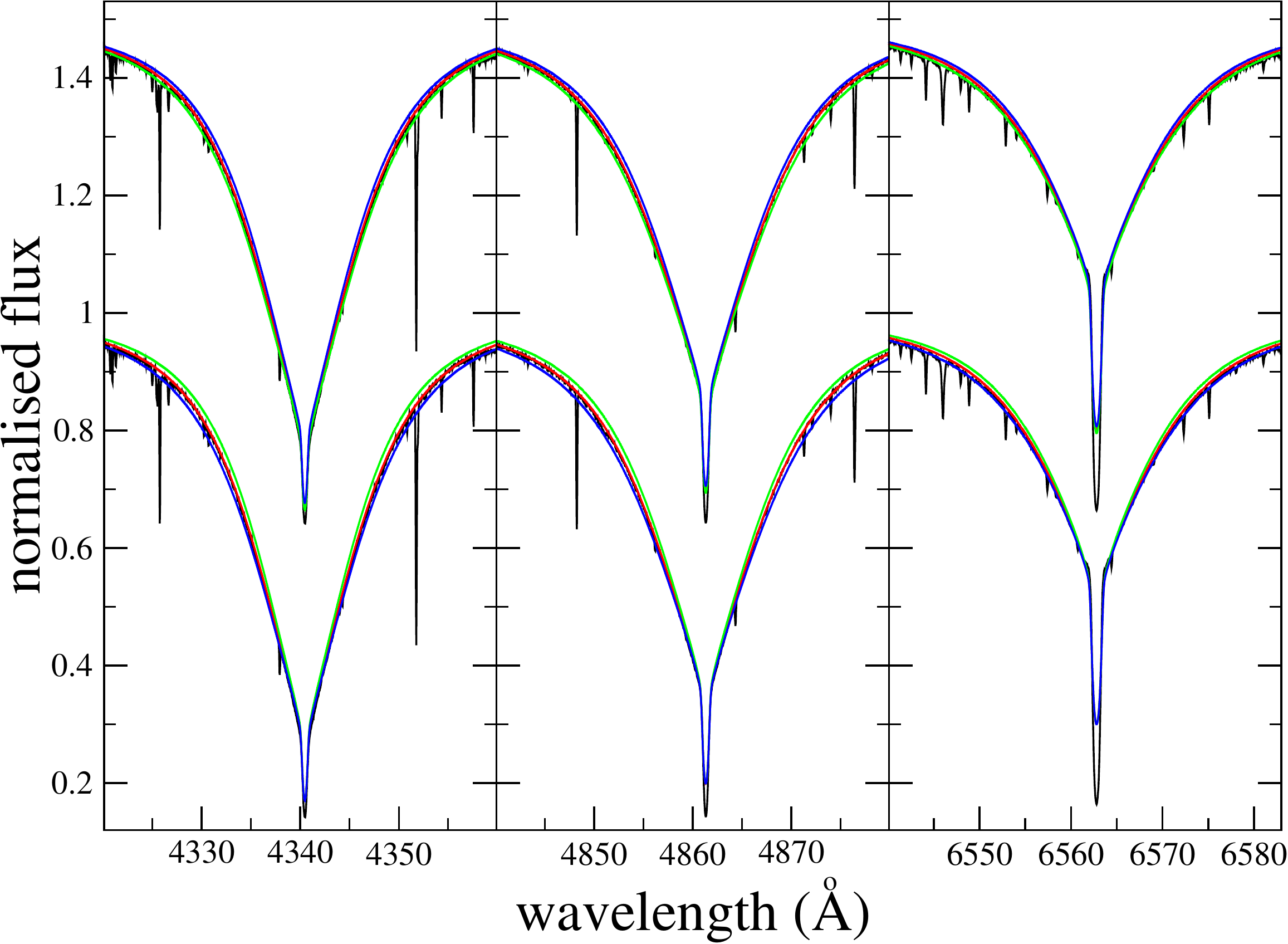}
\caption{Same as Fig.~\ref{hydrogen_21peg_hbeta}, but also for H$\gamma$ and
H$\alpha$.} 
\label{hydrogen_21peg} 
\end{center} 
\end{figure*}

\begin{figure*}[ht]
\begin{center}
\includegraphics[width=145mm]{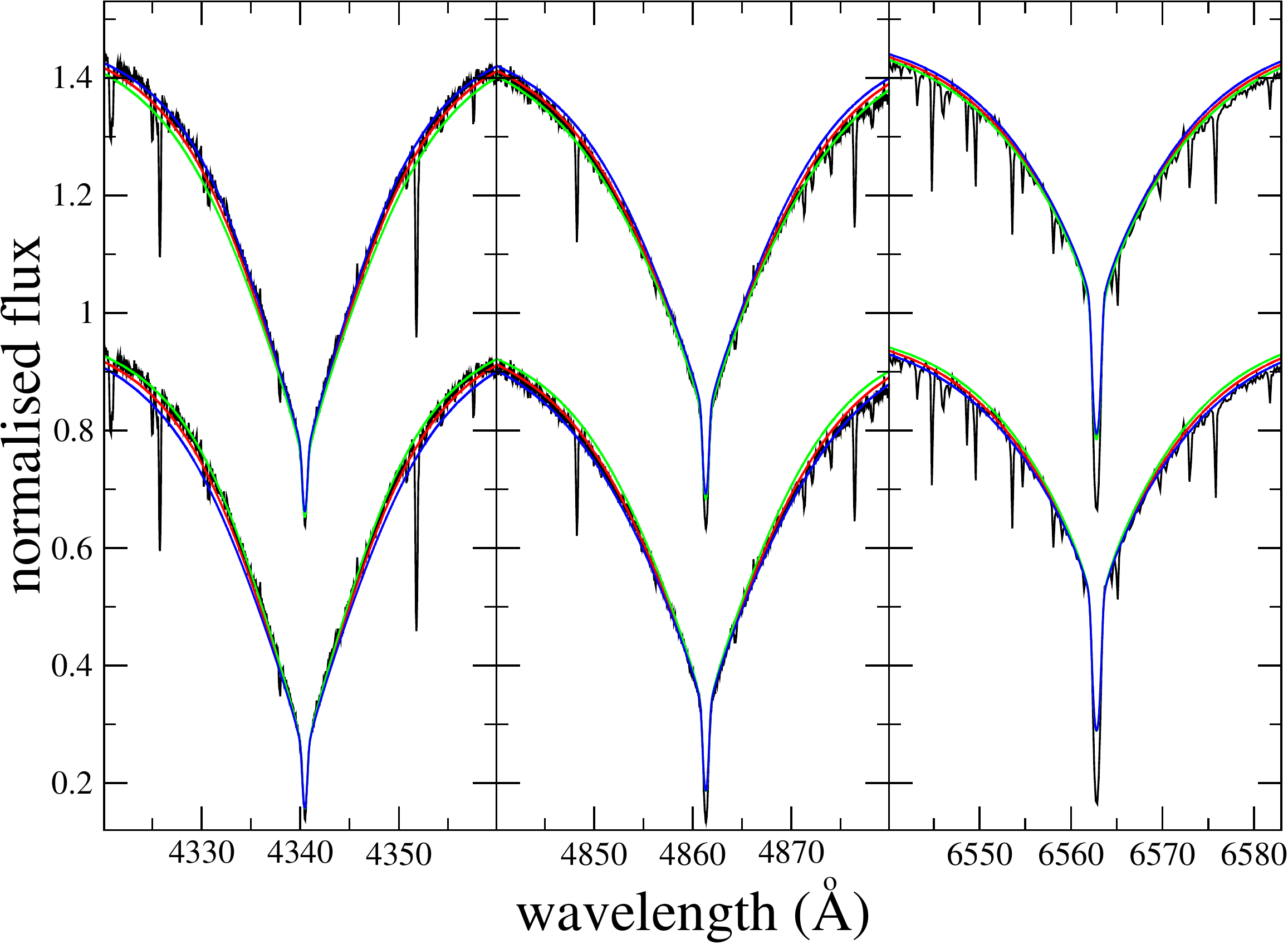}
\caption{Same as Fig.~\ref{hydrogen_21peg}, but for HD~145788.} 
\label{hydrogen_hd145788} 
\end{center} 
\end{figure*}

\begin{figure*}[ht]
\begin{center}
\includegraphics[width=145mm]{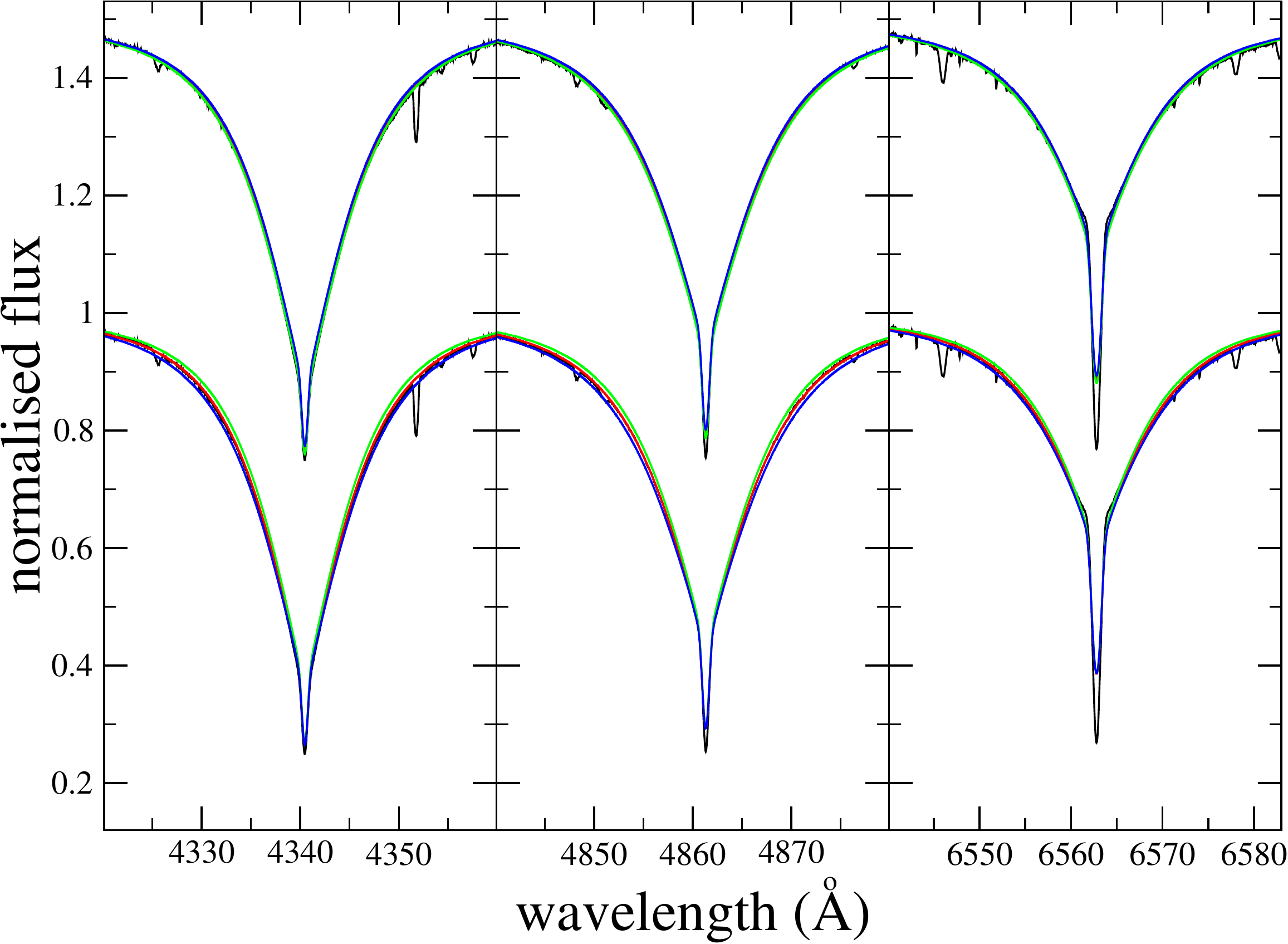}
\caption{Same as Fig.~\ref{hydrogen_21peg}, but for $\pi$~Cet.} 
\label{hydrogen_hd17081} 
\end{center} 
\end{figure*}

\begin{figure*}[ht]
\begin{center}
\rotatebox{90}{
\includegraphics[height=145mm]{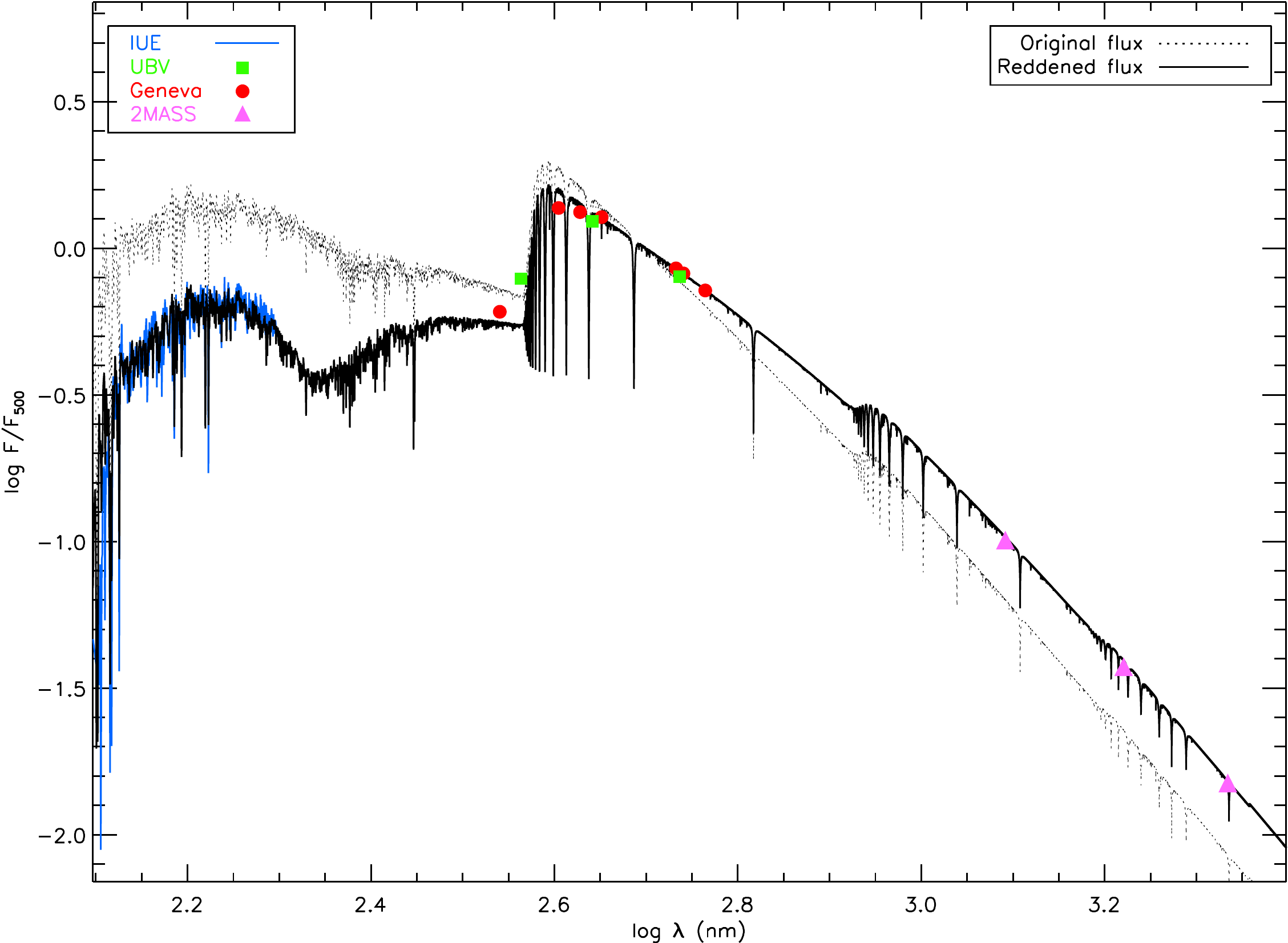}}
\caption{Comparison between \llm\ theoretical fluxes calculated with the 
fundamental parameters and abundances derived for HD~145788, taking into 
account a reddening of E(B-V)=0.20 (full black line) and without taking 
into account reddening (dashed black line), with IUE calibrated fluxes 
(full blue line), Johnson UBV photometry (green squares), Geneva photometry (red
circles) and 2MASS photmetry (violet triangles). The model fluxes were 
convolved to have approximately the same spectral resolution of the IUE fluxes 
({\it R}~$\sim$~900).}
\label{sph_hd145788} 
\end{center} 
\end{figure*}

In Fig.~\ref{alltosun} we show the derived abundances normalised to 
solar values \citep{met05}. In the following, we discuss the
stars of our sample individually.
%
\subsection{HD~145788}
HD~145788 shows a slight overabundance for almost all ions and typical Am 
abundance pattern for elements heavier than Ti. The overall overabundance and
the Am abundance pattern could be explained if HD~145788 was formed in a 
region of the sky with a metallicity higher than the solar region. To be able 
to check the possible Am classification of HD~145788, we can compare the 
obtained abundance pattern with the typical one for Am stars in clusters having 
enhanced metallicity, e.g. the Praesepe open cluster with an overall 
metallicity of [Fe/H]=0.14\,dex \citep{chen2003}. \citet{praesepe1} give the 
abundance pattern of eight Am stars belonging to the Praesepe cluster. All 
these stars show clear Am signatures: underabundances of CNO and Sc, and 
overabundances of the elements heavier than Ti. In HD~145788 the 
underabundances of CNO and Sc are not visible. For this reason we believe 
that the star cannot be classified as a hot Am star and that the observed 
abundance pattern stems from the composition of the cloud where HD~145788 
was formed.

Table~\ref{abundance} and Fig.~\ref{alltosun} show that with the adopted 
fundamental parameters of HD~145788 we get ionisation imbalance for the 
Fe-peak elements. With a small adjustment of the parameters within the 
error bars (\Teff: +100\,K, \logg: $-$0.05) it is possible to compensate for 
this imbalance for the Fe-peak elements, but the ionisation balance becomes 
worse for other elements such as C, Mg, and Ca. This is a clear sign that a 
\nlte\ analysis is needed for this object to understand whether the ionisation 
violation stems from \nlte\ effects or to some other physical effect. 
Unfortunately we cannot be completely certain of the adopted parameters 
because of the lack of spectrophotometric measurements.
\subsection{21~Peg}
The star 21~Peg shows solar abundances for almost all elements. At the 
temperature of 21~Peg, slowly rotating stars are usually hot Am, cool HgMn, 
or magnetic stars. 

The possibility of classifying 21~Peg as an Am star (which is potentially
suggested by the observed Sc underabundance) is excluded by the solar 
abundances of all the other mentioned indicators. As explained in 
Sect.~\ref{CaSc}, the \nlte\ correction for Ca is expected to be
positive, increasing the Ca abundance to the solar value, but detailed \nlte\ 
analysis should be performed for a more accurate determination. Almost nothing
is known about \nlte\ effects for Sc at these temperatures, so that to
understand whether the observed underabundance is real, a \nlte\ analysis
should be performed. 

\citet{adelman97}, \citet{adelman98}, \citet{adelman99}, and \citet{kocher03} 
have published abundances of several chemically normal, early-type stars. The 
derived Sc abundance for stars with an effective temperature similar to that 
of 21~Peg is almost always below the solar value, with nearly solar abundances 
for other Fe-peak elements. We thus conclude that the observed abundances of 
most elements but Ba in 21~Peg allow us to classify it as a normal early-type 
rather than Am star. Solar Mn abundance and the absence of 
\ion{Hg}{ii}~$\lambda$3984\,\AA\ line exclude any classification of 21~Peg as 
an HgMn star. 
\subsection{$\pi$~Cet}
The star $\pi$~Cet shows solar abundances for almost all elements. Only O, Ne, 
Na and Ar are clearly overabundant. For Ne or Na this most probably comes from 
\nlte\ effects. For argon, the \nlte\ effects are expected to be weak (see 
Sect.~\ref{NeAr}). In other chemically normal, early-type stars the Ar 
abundance appears to be above the solar one 
\citep[see][]{Lanz-Ar,praesepe1,adelman98}, leading to the conclusion that
indirect solar Ar abundance given by \citet{met05} is underestimated.

As for 21~Peg, the Sc underabundance of $\pi$~Cet is not an indication of a
possible Am peculiarity. The absence of magnetic field or of normal Mn 
abundance, including no trace of any Hg signature in the spectrum exclude a 
classification of the star as magnetic peculiar or non-magnetic HgMn object. 

The star $\pi$~Cet is a known binary with a period of about 7.5 years 
\citep{lacy} and is a Herbig AeBe star \citep{malfait}. The Herbig 
classification comes from a detected infrared excess at wavelengths longer 
than 10\,$\mu$m. The two spectra obtained with \espa\ show variability in the 
line profiles, small emission-like features close to the core of H$\alpha$, 
and emission features at the position of \ion{C}{i}~$\lambda\lambda$ $\sim$ 
8335 and 9405\,\AA\ in the near infrared. The width of \ion{C}{i} emission 
lines are exactly the same as expected for absorption line in $\pi$~Cet. The 
pre-main-sequence status of this star, which is very likely responsible for
these emissions, might also explain the variability observed in the 
spectral lines, as circumstellar absorption or emission coming from a 
proto-planetary disk. The variation in the line profile within one day excludes
the possibility that the observed changes are caused by the companion. 

Another explanation for the line variations comes from pulsation.
We performed a frequency analysis of the radial velocity  measurements given
by \citet{lacy} and the ones obtained from the two \espa\ spectra.
Preliminary results show that two frequencies appear in the amplitude spectrum:
one corresponding to the orbital period and  another one at 
$\sim$2.79\,day$^{-1}$. This frequency is consistent with the expected 
pulsation periods for SPB stars with effective temperature of $\pi$~Cet
\citep[see Fig.~5 of][]{pulsation}. In this way pulsation could explain the 
line-profile variation, while the presence of a disk 
around the star could explain the small emission visible in H$\alpha$. The 
\ion{C}{i} emission lines in the near infrared could be explained either 
by the disk or by \nlte\ effects \citep{emission}.

Only future photometric observations and time-resolved spectroscopy
and \nlte\ analysis could lead to a better understanding of the star's status.
\section{Conclusions: are solar abundances also a reference for early A-  
         and late B-type stars?}
One of the main goal of this work was to check whether the solar abundances can 
be taken as a reference for early-type stars. The same question has 
recently been discussed by \citet{P08} who analysed a sample of early B-type 
stars in the solar neighbourhood to compare the obtained \nlte\ abundances 
with the ones published by other authors for stars in the Orion nebula, 
various B-type stars, young F- and G-type stars, the interstellar medium, and 
the sun \citep{grevesse1996,met05}, and to check the chemical homogeneity of 
the solar neighbourhood. They obtained an excellent agreement between the 
\nlte\ abundances for He, C, N, Mg, Si, and Fe with the solar ones published by 
\citet{met05}, while the oxygen abundance lies between the solar values 
obtained by \citet{grevesse1996} and \citet{met05}, and the Ne abundance
is compatible with the one provided by \citet{grevesse1996}. 

The optical spectra of early B-type stars cannot provide reliable data for 
many other elements (Ca, Ti, Cr, Mn, Sr, Y, Zr), which are important for 
comparative abundance studies of chemically peculiar stars. While early A- 
and late B-type stars, investigated in the present paper, provide us 
with the abundances of up to 26 elements, most of which are based on enough 
spectral lines with accurately known atomic parameters.
Figure~\ref{alltosun} shows almost solar abundances for many elements in both 
21~Peg and $\pi$~Cet, while the observed abundance pattern in HD~145788 
gives a hint that the star may have been formed in a region of the sky at 
high metallicity.
In early-type stars it is possible to directly derive the He abundance, while
for the Sun it is only possible through astroseismological observations and
modelling. For this reason it is important to check that the He abundance is 
comparable to the solar value in several chemically normal early-type stars.

In the analysed stars, several elements (He, C, Al, S, V, Cr, Mn, Fe, Ni, Sr, 
Y, Zr) show abundances compatible with the revised solar data \citep{met05}, 
and when discrepancies are present they could be explained by \nlte\ effects 
(N, Na, Mg, Si, Ca, Ti, Nd). For Ne and Ar the expected \nlte\ corrections 
would lead to abundances close to those derived for early B-type stars 
\citep{Lanz-Ar,P08} or to the solar ones given by \citet{grevesse1996} 
instead of \citet{met05}. Non-LTE corrections were never calculated and 
should be determined for other elements that show differences with the 
solar abundance (P, Cl, Sc, Co). We found actual discrepancies with the solar 
abundance for oxygen in $\pi$~Cet and for Ba in 21~Peg. While the oxygen 
problem may be solved by careful \nlte\ analysis of all the available lines 
including the red and IR ones, the Ba overabundance cannot be explained by 
the current \nlte\ results.
The abundances obtained in this work for this set of three early B-type stars 
agree very well with the ones obtained by \citet{P08} for all the elements.

The published abundances of Ba in chemically normal early-type stars 
\citep{lemke90,adelman97,adelman99,kocher03} show a definite trend towards a 
Ba overabundance. The \nlte\ corrections for Ba should be positive, leading 
to an even greater discrepancy with the solar value, that probably does not 
represent early-type stars. 
   
Non-LTE effects are studied mainly in solar-type stars, low-metallicity stars, 
and giants, and in stars hotter than early B-type, where the effects are 
expected to be strong. Very few analyses have been performed for normal early 
A- and late B-type stars (e.g. Vega), and our study claims the real need of 
such analyses for many elements before making a definite conclusion about 
the solar abundances as standards for early-type stars.
\begin{acknowledgements}
This work is based on observations collected at the Nordic Optical Telescope 
(NOT) as part of programme number 35-001 and at the ESO 3.6\,m
telescope at Cerro La Silla (Chile). Part of this work is based on 
observations made with the Nordic Optical Telescope, operated on the island 
of La Palma jointly by Denmark, Finland, Iceland, Norway and Sweden, in 
the Spanish Observatorio del Roque de los Muchachos of the Instituto de 
Astrofisica de Canarias. This work is also based on observations
obtained at the Canada-France-Hawaii Telescope (CFHT), which is operated by the
National Research Council of Canada, the Institut National des Sciences de
l'Univers of the Centre National de la Rechereche Scientifique of France, and
the University of Hawaii. This work is supported by the
Austrian Science Foundation (FWF project P17890-N2 - LF, TR and OK), by
the Russian Foundation for Basic research (grant 08-02-00469a - TR)
and by the Presidium RAS Programme ''Origin and evolution of stars and
galaxies" (TR). GAW acknowledges support from the Academic Research Programme
(ARP) of the Department of National Defence (Canada). We thank 
D.~Lyashko for having developed and provided the reduction pipeline for the 
FIES data, V.~Tsymbal for providing us with an improved version of \width, and 
L.~Mashonkina for providing some \nlte\ estimates. We thank the anonymous
referee for the constructive comments. We thank A.~Ederoclite and 
L.~Monaco for the spectrum of HD~145788, and M.~Gruberbauer for the
frequency analysis. TR and LF thank D.~Shulyak for the fruitful
help, support and discussion of model atmospheres and spectral energy
distribution. This work made use of the MAST-IUE archive ({\tt
http://archive.stsci.edu/iue/}), of SAO/NASA ADS, SIMBAD, VIZIER and of
the VOSpec tool ({\tt http://www.euro-vo.org/pub/fc/software.html})
developed for the European Virtual Observatory. This publication makes use 
of data products from the Two Micron All Sky Survey, which is a joint project 
of the University of Massachusetts and the Infrared Processing and Analysis 
Center/California Institute of Technology, funded by the National Aeronautics 
and Space Administration and the National Science Foundation.
\end{acknowledgements}
\Online
%
\begin{figure*}[ht]
\sidecaption
\includegraphics[width=140mm]{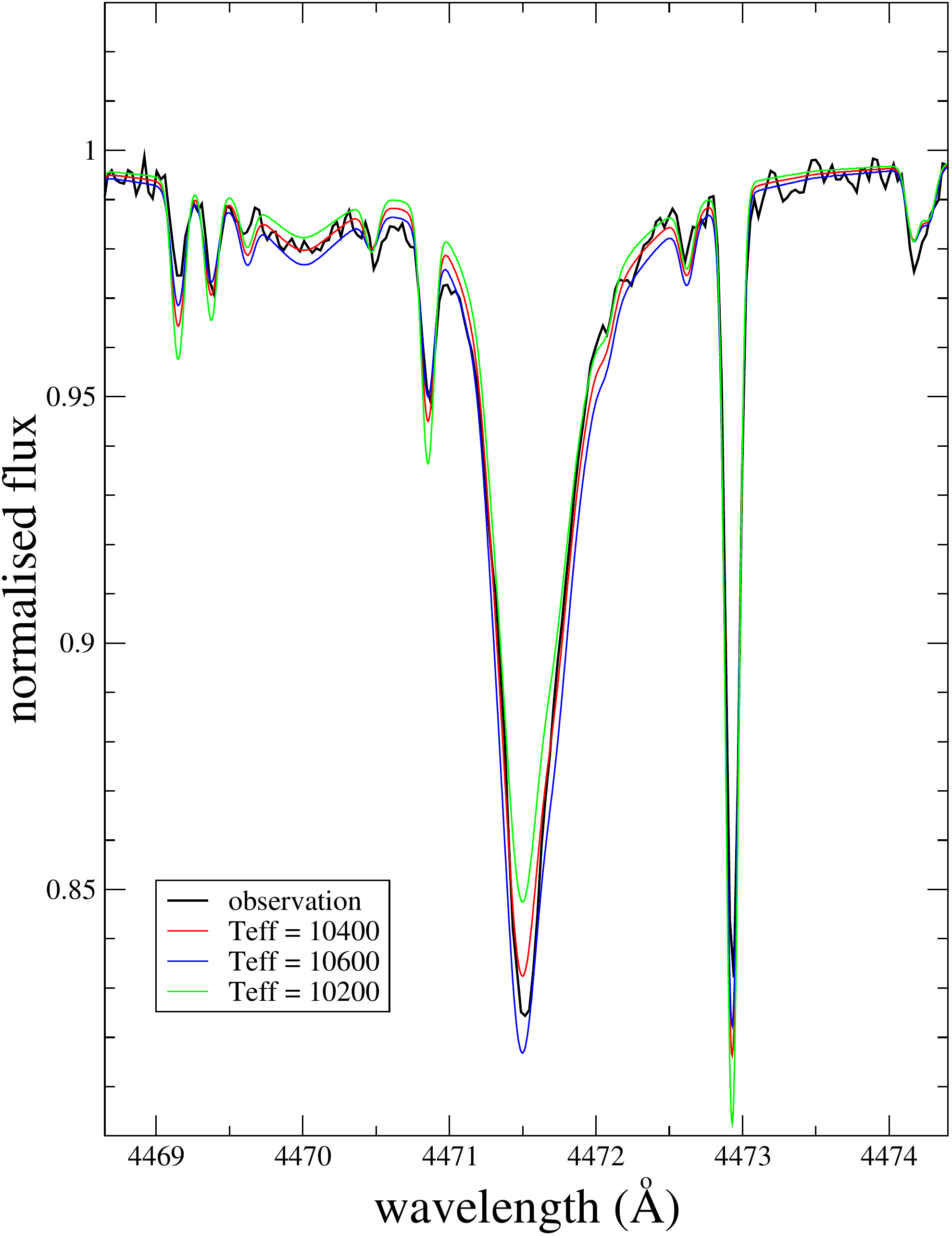}
\caption{Comparison between the observed spectrum of the \ion{He}{i} 
line at $\lambda\sim$4471\,\AA\ for 21~Peg (black thick line) and three 
synthetic profiles calculated with three different \Teff\ for the model 
atmosphere: 10200\,K (green line), the adopted 10400\,K (red line) and 
10600\,K (blue line).} 
\label{Hefig} 
\end{figure*}
\begin{table*}[ht]
\caption[ ]{Comparison of the derived abundances with previous determinations for 21~Peg.}
\label{comp_abn_21peg}
\begin{center}
\begin{tabular}{l|cc|cc|cc|cc}
\hline
\hline
Ion &\multicolumn{2}{|c|}{R08}&\multicolumn{2}{c|}{S81}      &\multicolumn{2}{c|}{S934}     &\multicolumn{2}{c}{D00}       \\
    &$\log (N/N_{\rm tot})$ & $n$ &$\log (N/N_{\rm tot})$ & $n$ &$\log (N/N_{\rm tot})$ & $n$ &$\log (N/N_{\rm tot})$  \\
\hline
\ion{He}{i }  & ~~$-$1.11$\pm$0.04 &  7 &                &   &                &   &  &  \\	   
\ion{C}{i}    & ~~$-$3.66$\pm$0.14 &  9 & -3.88          & 2 &                &   &  &  \\	   
\ion{C}{ii}   & ~~$-$3.65$\pm$0.05 &  4 &                &   &                &   &  &  \\	   
\ion{N}{i}    & ~~$-$3.95$\pm$0.12 &  4 &                &   &                &   &  &  \\	   
\ion{N}{ii}   & ~~$-$3.90:	        &  1 &                &   &                &   &  &  \\	   
\ion{O}{i}    & ~~$-$3.28$\pm$0.11 & 18 &                &   &                &   &  &  \\	   
\ion{Ne}{i}   & ~~$-$3.76$\pm$0.05 &  7 &                &   &                &   & -3.60$\pm$0.16 & 3 \\	   
              & ~~-3.89 (\nlte)    &  1 &                &   &                &   & -3.82 (\nlte) & 1 \\  
\ion{Na}{i}   & ~~$-$5.60:         &  1 &                &   &                &   &  &  \\	   
\ion{Mg}{i}   & ~~$-$4.42$\pm$0.12 &  5 &                &   &                &   &  &  \\	   
\ion{Mg}{ii}  & ~~$-$4.56$\pm$0.03 &  7 & -4.72$\pm$0.13 & 7 & -4.34$\pm$0.05 & 4 &  &  \\	   
\ion{Al}{i }  & ~~$-$5.89	         &  2 & -6.05          & 2 &                &   &  &  \\	   
\ion{Al}{ii}  & ~~$-$5.70$\pm$0.10 &  4 &                &   & -5.84          & 1 &  &  \\	   
\ion{Al}{iii} & 		                 &	   &                &   & -5.89$\pm$0.21 & 2 &  &  \\	   
\ion{Si}{i}   & ~~$-$4.95:         &  1 &                &   &                &   &  &  \\	   
\ion{Si}{ii}  & ~~$-$4.49$\pm$0.13 & 22 & -4.85          & 2 & -4.63$\pm$0.07 & 2 &  &  \\	   
\ion{Si}{iii} & ~~$-$4.26:         &  2 &                &   &                &   &  &  \\	   
\ion{P}{ii}   & ~~$-$6.37$\pm$0.06 &  3 &                &   &                &   &  &  \\	   
\ion{S}{ii}   & ~~$-$4.86$\pm$0.13 & 26 &                &   &                &   &  &  \\	   
\ion{Ca}{i}   & ~~$-$5.84$\pm$0.11 &  3 & -5.76          & 1 &                &   &  &  \\	   
\ion{Ca}{ii}  & ~~$-$5.98$\pm$0.08 &  5 & -6.24          & 1 &                &   &  &  \\	   
\ion{Sc}{ii}  & ~~$-$9.37$\pm$0.10 &  7 & -9.47          & 3 &                &   &  &  \\	   
\ion{Ti}{ii}  & ~~$-$7.23$\pm$0.09 & 59 & -7.15$\pm$0.16 &26 &                &   &  &  \\	   
\ion{V}{ii}   & ~~$-$7.98$\pm$0.06 &  9 & -7.78$\pm$0.16 & 6 &                &   &  &  \\	   
\ion{Cr}{i}   & ~~$-$6.29$\pm$0.09 &  5 &                &   &                &   &  &  \\	   
\ion{Cr}{ii}  & ~~$-$6.20$\pm$0.10 & 68 & -6.39$\pm$0.25 &17 & -6.30$\pm$0.01 & 5 &  &  \\	   
\ion{Mn}{i}   & ~~$-$6.54$\pm$0.21 &  5 & -6.08          & 2 &                &   &  &  \\	   
\ion{Mn}{ii}  & ~~$-$6.51$\pm$0.17 & 19 &                &   & -6.39$\pm$0.10 & 3 &  &  \\	   
\ion{Fe}{i}   & ~~$-$4.52$\pm$0.13 &108 & -4.80$\pm$0.21 &32 &                &   &  &  \\	   
\ion{Fe}{ii}  & ~~$-$4.50$\pm$0.12 &406 & -4.79$\pm$0.18 &23 & -4.59$\pm$0.05 &10 &  &  \\	   
\ion{Fe}{iii} & ~~$-$4.60$\pm$0.06 &  3 &                &   &                &   &  &  \\	   
\ion{Co}{ii}  & ~~$-$6.75$\pm$0.18 &  3 &                &   & -6.01$\pm$0.01 & 3 &  &  \\	   
\ion{Ni}{i}   & ~~$-$5.71$\pm$0.05 & 10 &                &   &                &   &  &  \\	   
\ion{Ni}{ii}  & ~~$-$5.61$\pm$0.09 & 23 & -6.13          & 5 & -5.70$\pm$0.02 & 4 &  &  \\	   
\ion{Zn}{i}   & ~~$-$6.96:         &  1 &                &   & -7.29$\pm$0.11 & 3 &  &  \\	   
\ion{Sr}{ii}  & ~~$-$9.10:         &  2 & -9.18          & 2 &                &   &  &  \\	   
\ion{Y}{ii}   & ~~$-$9.76$\pm$0.15 &  4 & -9.27          & 1 &                &   &  &  \\	   
\ion{Zr}{ii}  & ~~$-$9.48$\pm$0.28 &  4 & -9.09          & 1 &                &   &  &  \\	   
\ion{Ba}{ii}  & ~~$-$9.19$\pm$0.06 &  3 &                &   &                &   &  &  \\	   
\ion{Nd}{iii} & ~$-$10.09$\pm$0.07 &  3 &                &   &                &   &  &  \\	   
\hline											  
\end{tabular}
\end{center}
\smallskip

In column~2, the meaning of a colon is the same as in Table~\ref{abundance}.\\ \\
R08: this work; S81: \citet{sadakane81}; 
S934: \citet{smith93,smith1993,smith1994}; D00: \citet{Dworetsky00}.
\end{table*}

\begin{table*}[ht]
\caption[ ]{Comparison of the derived abundances with previous dterminations for $\pi$~Cet.}
\label{comp_abn_picet}
\begin{center}
\begin{tabular}{l|cc|cc|cc|cc|cc}
\hline
\hline
Ion &\multicolumn{2}{|c|}{R08}&\multicolumn{2}{c|}{A91}            &\multicolumn{2}{c|}{S934}     & \multicolumn{2}{c|}{D00}     & \multicolumn{2}{c}{A04}      \\
    &$\log (N/N_{\rm tot})$ & $n$ &$\log (N/N_{\rm tot})$ & $n$ &$\log (N/N_{\rm tot})$ & $n$ &$\log (N/N_{\rm tot})$ & $n$ &$\log (N/N_{\rm tot})$  & $n$ \\
\hline
\ion{He}{i}   & -0.97$\pm$0.04 &  6 & -1.11$\pm$0.06 &  6 &	       &   & 	          &   & 	     &   \\   
\ion{C}{ii}   & -3.58$\pm$0.07 &  7 & -3.81$\pm$0.07 &  4 &	       &   & 	          &   & 	     &   \\   
\ion{N}{i}    & -4.03$\pm$0.13 & 10 &		   &	&              &   & 	          &   & 	     &   \\   
\ion{N}{ii}   & -3.74$\pm$0.07 &  9 & -3.92$\pm$0.15 &  3 &	       &   & 	          &   & 	     &   \\   
\ion{O}{i}    & -3.06$\pm$0.14 &  9 & -3.34        &  2 &	       &   & 	          &   & -3.29$\pm$0.03 & 4 \\
\ion{O}{ii}   & -3.04:         &  2 &		   &	&              &   & 	          &   & 	     &   \\   
\ion{Ne}{i}   & -3.66$\pm$0.09 & 20 &		   &	&              &   & -3.71$\pm$0.19 & 6 & 	     &   \\   
              & -3.86 (NLTE)   &  1 &		   &	&              &   & -3.86 (NLTE) & 1 & 	     &   \\   
\ion{Na}{i}   & -5.23$\pm$0.07 &  3 &		   &	&              &   & 	          &   & 	     &   \\   
\ion{Mg}{i}   & -4.27:         &  2 & -4.88        &  1 &	       &   & 	          &   & 	     &   \\   
\ion{Mg}{ii}  & -4.47$\pm$0.16 & 10 & -4.56$\pm$0.08 &  7 & -4.28$\pm$0.05 & 4 & 	          &   & -4.57$\pm$0.02 & 4 \\
\ion{Al}{i}   & -5.57:         &  2 & -5.85        &  2 &	       &   & 	          &   & 	     &   \\   
\ion{Al}{ii}  & -5.73$\pm$0.27 &  8 &		   &    & -6.04        & 1 & 	          &   & 	     &   \\   
\ion{Al}{iii} & -5.30$\pm$0.02 &  3 & -5.36        &  1 & -5.79$\pm$0.10 & 2 & 	          &   & 	     &   \\   
\ion{Si}{i}   & -4.80:         &  1 &		   &	&              &   & 	          &   & 	     &   \\   
\ion{Si}{ii}  & -4.41$\pm$0.20 & 31 & -4.56$\pm$0.12 &  5 & -4.44$\pm$0.05 & 2 & 	          &   & -4.44$\pm$0.28 & 4 \\
\ion{Si}{iii} & -4.16:         &  2 & -4.99        &  1 &              &   &	          &   & 	     &   \\    
\ion{P}{ii}   & -6.38$\pm$0.19 &  9 &		   &    &              &   &	          &   & 	     &   \\    
\ion{P}{iii}  & -6.19:         &  1 &		   &    &              &   &	          &   & 	     &   \\    
\ion{S}{ii}   & -4.78$\pm$0.16 & 31 & -4.86$\pm$0.18 & 18 &              &   &	          &   & -4.90	     & 1 \\
\ion{Cl}{ii}  & -6.95:         &  2 &		   &    &              &   &	          &   & 	     &   \\    
\ion{Ar}{i}   & -4.86:         &  2 &		   &    &              &   &	          &   & 	     &   \\    
\ion{Ar}{ii}  & -5.24$\pm$0.19 &  6 &		   &    &              &   &	          &   & 	     &   \\    
\ion{Ca}{ii}  & -5.77:         &  2 & -5.72        &  1 &              &   &	          &   & 	     &   \\    
\ion{Sc}{ii}  & -9.31:         &  1 &		   &    &              &   &	          &   & 	     &   \\    
\ion{Ti}{ii}  & -7.42$\pm$0.08 & 11 & -7.17$\pm$0.24 & 14 &              &   &	          &   & -7.23$\pm$0.13 & 7 \\
\ion{Cr}{ii}  & -6.41$\pm$0.10 & 21 & -6.58$\pm$0.19 & 15 & -6.00$\pm$0.2  & 5 & 	          &   & -6.58$\pm$0.08 &10 \\
\ion{Mn}{ii}  & -6.50$\pm$0.09 &  3 &		   &    & -6.44$\pm$0.05 & 3 & 	          &   & 	     &   \\   
\ion{Fe}{i}   & -4.53$\pm$0.22 &  7 &		   &    &	       &   & 	          &   & 	     &   \\   
\ion{Fe}{ii}  & -4.58$\pm$0.14 &186 & -4.66$\pm$0.20 & 59 & -4.55$\pm$0.05 &10 & 	          &   & -4.78$\pm$0.19 &29 \\ 
\ion{Fe}{iii} & -4.52$\pm$0.10 &  4 & -4.82        &  1 &	       &   & 	          &   & 	     &   \\   
\ion{Co}{ii}  & -6.93:         &  1 &		   &    & -6.30$\pm$0.1  & 3 & 	          &   & 	     &   \\   
\ion{Ni}{ii}  & -5.76$\pm$0.19 & 17 & -6.02        &  3 & -5.80$\pm$0.2  & 4 & 	          &   & -5.89$\pm$0.02 & 2 \\
\ion{Sr}{ii}  & -9.15:         &  2 &		   &    &	       &   & 	          &   & 	     &   \\   
\hline
 & \multicolumn{2}{|c|}{\vmic\ = 1 \kms} & \multicolumn{2}{c|}{\vmic\ = 0 \kms} & \multicolumn{2}{c|}{\vmic\ = 0 \kms} & & & \multicolumn{2}{c}{\vmic\ = 3 \kms} \\
\hline
\end{tabular}
\end{center}
\smallskip

In column~2, the meaning of a colon is the same as in Table~\ref{abundance}.\\ \\
R08: this work; A91: \citet{adelman91}; 
S934: \citet{smith93,smith1993,smith1994}; D00: \citet{Dworetsky00};
A04: \citet{acke04}
\end{table*}

%
\begin{table*}[ht]
\label{Si2-gf}
\caption{A collection of the experimental and theoretical transition probabilities and Stark 
widths for the observed \ion{Si}{ii} lines. Errors are given in parenthesis.}
\begin{footnotesize}
\begin{center}
\begin{tabular}{lr|r|r|r|r|r|r|r|r|cc}
\hline
\hline
Wavelength & \exc &        & \multicolumn{7}{|c|}{\loggf} &\multicolumn{2}{|c}{$\log\,\gamma_{\rm St}$}\\
\AA        & eV   &   NIST & Barach   &     SG   &    BBCB  &    BBC   &   Math   &    Wilke  &  AJPP&Wilke & LDA  \\
\hline   
3853.665 &  6.858 & -1.341 &          &-1.32(06) &-1.52     &-1.39(06) &-1.28(05) &-1.53(11)  &-1.44 &-5.12 &-5.15 \\
3856.018 &  6.859 & -0.406 &          &-0.34(06) &-0.56     &-0.43(05) &-0.36(06) &-0.65(10)  &-0.49 &-5.15 &-5.15 \\
3862.595 &  6.858 & -0.757 &          &-0.83(06) &-0.82     &-0.80(06) &-0.80(07) &-0.92(10)  &-0.75 &-5.14 &-5.15 \\
4072.709 &  9.837 & -2.700 &          &-2.37(05) &	         &	         &-2.70(10) &-2.40(16)  &-1.89 &-4.89 &-4.76 \\
4075.452 &  9.839 & -1.403 &          &-1.40(05) &          &          &-1.40(10) &-1.64(16)  &-0.95 &-4.89 &-4.76 \\
4076.780 &  9.837 & -1.700 &          &-1.67(05) &	         &	         &-1.70(10) &-1.75(16)  &-1.20 &-4.89 &-4.76 \\
4128.054 &  9.837 &  0.359 &          & 0.45(04) & 0.32     &          & 0.36(12) & 0.22(08)  & 0.38 &-4.92 &-4.88 \\
4130.872 &  9.839 & -0.783 &          &          &-0.82     &          &          &           &-0.77 &      &-4.88 \\
4130.894 &  9.839 &  0.552 &          & 0.50(04) & 0.48    	&         	& 0.55(12) & 0.38(08)  & 0.53 &-4.93 &-4.88 \\
4190.707 & 13.492 &        &          &          &	         &	         &-0.17(10) &-0.33(11)  &-0.35 &-5.25 &-5.26 \\
4198.134 & 13.487 &        &          &          &	         &	         &-1.46(11) &-0.60(11)  &-0.61 &-5.26 &-5.26 \\
4200.658 & 12.525 & -0.889 &          &          &	         &	         &          &           &      &      &-3.43 \\
4621.418 & 12.525 & -0.608 &          &          &	         &	         &          &           &      &      &-3.69 \\
4621.722 & 12.526 & -0.453 &          &          &	         &	         &          &           &      &      &-3.69 \\
5041.023 & 10.067 &  0.029 &          & 0.29(03) &          & 0.09(04) & 0.03(07) & 0.18(03)  & 0.19 &-4.84 &-4.78 \\
5055.983 & 10.074 &  0.523 &          & 0.59(03) &          & 0.46(04) & 0.52(08) & 0.48(03)  & 0.44 &-4.79 &-4.78 \\
5056.316 & 10.074 & -0.492 &          &-0.36(03) &          &-0.49(04) &          &           &-0.51 &-4.79 &-4.78 \\
5466.460 & 12.525 & -0.237 &          &          &	         &	         &          &           &      &      &-3.99 \\
5469.451 & 12.880 & -0.762 &          &          &	         &	         &          &           &      &      &      \\
5540.807 & 14.489 &        &-0.83(08) &          &          &          &-1.20(10) &           &-0.80 &      &      \\
5576.661 & 14.504 &        &-0.47(08) &          &	         &	         &-0.47(10) &-0.62(13)  &-0.49 &-5.46 &      \\
5632.966 & 14.186 & -0.818 &-0.62(08) &          &          &-0.82(06) &          &-0.82(09)  &-0.69 &-5.44 &      \\
5639.477 & 14.528 &        &-0.07(06) &          &          &-0.30(07) &-0.07(12) &-0.18(09)  &-0.32 &-5.44 &      \\
5669.563 & 14.200 &  0.286 & 0.32(06) &          &          & 0.28(07) &          & 0.12(08)  & 0.25 &-5.53 &      \\
5688.817 & 14.186 &  0.126 & 0.08(06) &          &	         &	0.13(06) &          & 0.00(09)  & 0.08 &-5.50 &      \\
5701.370 & 14.175 & -0.057 &-0.10(06) &          &          &-0.06(07) &          &-0.28(09)  &-0.10 &-5.53 &      \\
5800.454 & 14.504 &        &-0.04(08) &          &          &-0.16(07) &-0.12(12) &-0.16(11)  &-0.17 &-5.37 &      \\
5806.731 & 14.489 &        &-0.10(08) &          &          &-0.13(07) &-0.11(12) &-0.18(10)  &-0.20 &-5.47 &      \\
5827.752 & 14.489 &        &-0.91(18) &          &          &-0.79(07) &-1.00(10) &-1.16(11)  &-0.90 &-5.45 &      \\
5846.121 & 14.504 &        &-0.51(08) &          &          &-0.29(07) &-0.57(12) &-0.33(14)  &-0.70 &-5.35 &      \\
5867.480 & 14.504 &        &          &          &	         &	         &          &           &-0.20 &      &      \\
5868.443 & 14.528 &        & 0.42(06) &          &	         &	         & 0.40(14) & 0.50(11)  & 0.20 &-5.36 &      \\
5957.559 & 10.067 & -0.225 &          &-0.30(02) &          &-0.22(03) &          &-0.29(06)  &-0.33 &-5.02 &-4.84 \\
5978.930 & 10.074 &  0.084 &          & 0.00(02) &          & 0.08(03) &          & 0.03(06)  &-0.03 &-5.01 &-4.84 \\
6347.109 &  8.121 &  0.149 &          & 0.18(05) & 0.30     & 0.20(04) & 0.12(07) & 0.29(08)  & 0.17 &-5.31 &-5.05 \\
6371.371 &  8.121 & -0.082 &          &-0.06(05) &-0.00     &-0.03(03) &-0.04(07) &-0.02(08)  &-0.13 &-5.32 &-5.05 \\
6660.532 & 14.504 &  0.162 &          &          &	         &	         &          & 0.23(09)  & 0.24 &-5.51 &      \\
6665.030 & 14.495 & -0.240 &          &          &	         &	         &          &-0.18(09)  &-0.16 &-5.54 &      \\
6671.841 & 14.528 &  0.409 &          &          &	         &	0.41(07) &          & 0.46(09)  & 0.52 &-5.58 &      \\
6699.431 & 14.495 & -0.247 &          &          &          &-0.25(08) &          &           &-0.16 &      &      \\
7848.816 & 12.525 &  0.316 &          &          &	         &	         &          &           & 0.34 &      &-4.34 \\
7849.618 & 12.526 & -0.831 &          &          &	         &	         &          &           &-0.80 &      &-4.34 \\
7849.722 & 12.526 &  0.470 &          &          &	         &	         &          &           & 0.50 &      &-4.34 \\
\hline   
\end{tabular}
\end{center}
\end{footnotesize}
\smallskip

NIST -- \citet{KP08} -- critical compilation; Barach -- \citet{Barach} -- pulsed arc; SG -- \citet{SG69} -- arc emission spectra;
BBCB -- \citet{BBCB} -- beam-foil lifetime measurements and theortical branching ratios; BBC -- \citet{BBC95} -- laser-induced plasma; 
Math -- \citet{Math01} -- laser-induced plasma; Wilke -- \citet{Wilke03} -- laser-induced plasma, Stark widths, shifts and transition probabilities;
AJPP -- \cite{AJPP81}  -- theoretical calculations; LDA -- \citet{Si2-Stark88} -- Stark widths, semi-classical calculations.
\end{table*}

%
%
\longtab{9}{
\begin{longtable}[]{ccl|cc|cc|cc|r}
\caption{\label{line_abund}Linelist of the lines used for the abundance analysis.}\\
\hline\hline \\
Element    &      &          & \multicolumn{2}{|c|}{HD~145788} & \multicolumn{2}{c|}{21~Peg} & \multicolumn{2}{c|}{$\pi$~Cet} &            \\
Wavelength & \exc & \loggf~~ & EQW  & abundance                & EQW  & abundance		   & EQW  & abundance		    & Ref \loggf \\
\AA        & eV   &          & m\AA & dex                      & m\AA & dex 		   & m\AA & dex			    &            \\
\hline   
\endfirsthead
\caption{continued.}\\
\hline\hline \\
Element    &      &          & \multicolumn{2}{|c|}{HD~145788} & \multicolumn{2}{c|}{21~Peg} & \multicolumn{2}{c|}{$\pi$~Cet} &            \\
Wavelength & \exc & \loggf~~ & EQW  & abundance                & EQW  & abundance		   & EQW  & abundance		    & Ref \loggf \\
\AA        & eV   &          & m\AA & dex                      & m\AA & dex 		   & m\AA & dex			    &            \\
\hline     
\endhead
\hline
\endfoot
\hline
\hline
\endlastfoot
\ion{He}{i} & & & & & & & & & \\
4026.1850 & 20.964 & -2.620(005) & S     & -1.13 &  S    & -1.15 &       &       &WSG  \\ 
4026.1870 & 20.964 & -1.450(005) & S     &	      &  S    &       &       &       &WSG  \\  
4026.1870 & 20.964 & -0.700(005) & S     &	      &  S    &       &       &       &WSG  \\  
4026.1980 & 20.964 & -1.450(005) & S     &	      &  S    &       &       &       &WSG  \\  
4026.3580 & 20.964 & -1.320(005) & S     &	      &  S    &       &       &       &WSG  \\  
4026.1990 & 20.964 & -0.970(005) & S     &	      &  S    &       &       &       &WSG  \\  
4026.3580 & 20.964 & -1.320(005) & S     &	      &  S    &       &       &       &WSG  \\
\hline
\end{longtable}
\smallskip

Wavelengths and excitation potentials are taken from the VALD database. The adopted \loggf\ values 
are taken from different sources which are listed in the last column. Errors in 
\loggf\ values if available are given in paranthesis. ''S" means that the line abundance was 
determined by fitting an observed line profile and not with equivalent width.\\ \\
AJPP81 - \citet{AJPP81};\\
BBPL - \citet{BBPL};\\ 
BGF - \citet{BGF};\\ 
BHN - \citet{BHN};\\
BK  - \citet{BK};\\ 
BKK - \citet{BKK};\\
BL - \citet{BL91};\\
BM - \citet{BM};\\
BWL - \citet{BWL};\\ 
FMW - \citet{FMW};\\ 
H - \citet{H}; \\
HL - \citet{HL};\\ 
K88 - \citet{K88}, and KurXX - Kurucz calculations of the XX year;\\ 
KG - \citet{KG};\\ 
KP - \citet{KP};\\ 
KP08 - \citet{KP08};\\ 
KSG - \citet{KSG};\\ 
L - \citet{L};\\
LD - \citet{LD};\\ 
LNAJ - \citet{LNAJ};\\ 
MFW - \citet{MFW};\\ 
MRB - \citet{MRB};\\ 
Mult - estimated from multiplet table intensity;\\
MW - \citet{MW};\\  
MWRB - \citet{MWRB};\\ 
NIST08 - \citet{NIST08};\\ 
PTP - \citet{PTP};\\ 
RHL - \citet{RHL};\\
RPU - \citet{RPU};\\ 
RRKB - \citet{RRKB};\\ 
RU - Raassen \& Uylings database (ftp://ftp.wins.uva.nl/pub/orth);\\
S - \citet{S};\\ 
SG - \citet{SG};\\
SN - \citet{SN};\\
SR - \citet{SR};\\
SW - \citet{SW};\\
T - \citet{T};\\
TB - TopBase, \citet{TB};\\
Wa - \citet{W};\\ 
Wb - \citet{War};\\
WF - \citet{WF};\\ 
WLa - \citet{WLa};\\ 
WSG - \citet{WSG};\\ 
WSM - \citet{WSM}. \\
}

\end{document}